\documentclass[usenatbib]{mn2e}
\usepackage{epsfig}
\usepackage{amsmath,amssymb}
\usepackage{aas_macros}

\usepackage[dvips]{color}
\definecolor{rev}{rgb}{1.0,0.0,0.0}
\definecolor{rev2}{rgb}{0.0,0.0,1.0}
\definecolor{cut}{rgb}{1.0,0.547,0.0}

\newcommand{\Caf}{SDSS J102915+172927}
\newcommand{\percc}{{\rm cm^{-3}}}
\newcommand{\um}{{\rm \mu m}}
\newcommand{\E}[1]{\times 10^{#1}}
\newcommand{\dex}{{\rm dex}}

\newcommand{\Mg}{{\rm Mg}}

\newcommand{\Si}{{\rm Si}}

\newcommand{\Pyroxene}{{\rm MgSiO_3}}
\newcommand{\Olivine}{{\rm Mg_2SiO_4}}
\newcommand{\Magnetite}{{\rm Fe_3O_4}}
\newcommand{\Silica}{{\rm SiO_2}}
\newcommand{\Troilite}{{\rm FeS}}
\newcommand{\Alumina}{{\rm Al_2O_3}}
\newcommand{\Magnesia}{{\rm MgO}}

\newcommand{\nH}{n_{{\rm H}}}
\newcommand{\mH}{m_{{\rm H}}}

\newcommand{\namb}{n_{{\rm amb}}}
\newcommand{\rhoamb}{\rho _{{\rm amb}}}

\newcommand{\XH}{X_{{\rm H}}}
\newcommand{\Zsun}{{\rm Z_{\bigodot}}}
\newcommand{\Msun}{\rm M_{\bigodot}}

\newcommand{\Zcrit}{Z_{{\rm cr}}}
\newcommand{\Mpr}{M_{{\rm pr}}}
\newcommand{\fdep}{f_{{\rm dep}}}
\newcommand{\fdepini}{f_{{\rm dep},0}}

\newcommand{\Noz}{N07}
\newcommand{\Sch}{S12}

\title[SN dust and grain growth: The critical metallicity]
{Supernova dust formation and the grain growth in the early universe:
The critical metallicity for low-mass star formation}

\author[G. Chiaki et al.]
{Gen Chiaki,$^{1}$\thanks{E-mail: gen.chiaki@utap.phys.s.u-tokyo.ac.jp}
Stefania Marassi,$^{2}$
Takaya Nozawa,$^{3}$ 
Naoki Yoshida,$^{1,4}$
\newauthor 
Raffaella Schneider,$^{2}$
Kazuyuki Omukai,$^{5}$
Marco Limongi,$^{2,4,6}$ and
Alessandro Chieffi$^{7}$
\\
$^{1}$Department of Physics, Graduate School of Science, University of Tokyo, 
7-3-1 Hongo, Bunkyo, Tokyo 113-0033, Japan \\
$^{2}$INAF/Osservatorio Astronomico di Roma, via Frascati 33, 00040 Roma, Italy \\
$^{3}$National Astronomical Observatory of Japan, Mitaka, Tokyo
181-8588, Japan \\
$^{4}$Kavli Institute for the Physics and Mathematics of the Universe (WPI), 
Todai Institutes for Advanced Study, \\
The University of Tokyo, Kashiwa, Chiba 277-8583, Japan \\
$^{5}$Astronomical Institute, Tohoku University, 6-3 Aramaki, Aoba, Sendai 980-8578, Japan \\
$^{6}$European Southern Observatory, Karl-Schwarzschild-Str. 2, 85748 Garching bei Munchen, Germany \\
$^{7}$INAF/IASF, Via Fosso del Cavaliere 100, 00133 Roma, Italy}

\begin{document}

\date{}

\pagerange{\pageref{firstpage}--\pageref{lastpage}} \pubyear{2014}

\maketitle

\label{firstpage}

\begin{abstract}
We investigate the condition for the formation of 
low-mass second-generation stars in the early universe.
It has been proposed that gas cooling by dust thermal
emission can trigger fragmentation of a low-metallicity star-forming 
gas cloud.
In order to determine 
the critical condition in which dust cooling induces
the formation of low-mass stars,
we follow the thermal evolution of a collapsing cloud
by a one-zone semi-analytic collapse model.
Earlier studies
assume the dust amount in the local universe, where all refractory 
elements are depleted onto grains, and/or
assume the constant dust amount during gas collapse. 
In this paper, we employ the models of dust formation and destruction in early supernovae
to derive the realistic dust compositions and size distributions for multiple species
as the initial conditions of our collapse calculations.
We also follow accretion of heavy elements in the gas phase onto dust grains,
i.e., grain growth, during gas contraction.
We find that grain growth well alters the fragmentation property of the clouds.
The critical conditions can be written by the gas metallicity $Z_{\rm cr}$ and
the initial depletion efficiency $\fdepini$ of gas-phase metal onto grains, or dust-to-metal mass ratio, as 
$\left (Z_{\rm cr}/ 10^{-5.5} \ \Zsun \right) = \left( \fdepini /0.18 \right) ^{-0.44}$
with small scatters in the range of $Z_{\rm cr} = [0.06$--$3.2]\E{-5} \ \Zsun$.
We also show that the initial dust composition and size distribution are important to determine $\Zcrit$.
\end{abstract}


%%%%%%%%%%%%%%%%%%%%%%%%%%%%%%%%%%%%%%%%%%%%%
% INTRODUCTION %%%%%%%%%%%%%%%%%%%%%%%%%%%%%%%%%%%
%%%%%%%%%%%%%%%%%%%%%%%%%%%%%%%%%%%%%%%%%%%%%

\section{INTRODUCTION}

The first stars (Population III or Pop III stars) 
formed in metal-free gas are thought to be
predominantly massive with several tens to thousand solar masses
\citep{Bromm01, Abel02, Omukai03, Yoshida06, Hosokawa11, Susa14, Hirano14}.
This is in stark contrast with the typical mass of the Galactic stars 
which is less than the solar mass \citep{Kroupa02}.
Therefore, how and when the transition of the typical stellar mass occurred
is one of the critical issues for understanding the star forming history
throughout the cosmic time. The long-lived low-mass stars discovered in
the Galactic halo may be the fossils of the first low-mass stars.

Recent studies propose that the formation of the first low-mass stars is driven
by the fine-structure line cooling by carbon and oxygen \citep{Bromm01, BrommLoeb03,
SantoroShull06, Frebel07, Ji14}.
However, the gas density at which the fine-structure line cooling can cause 
fragmentation of star-forming clouds is 
$\nH \sim 10^4$--$10^5 \ \percc$, where the typical mass
of the fragments (comparable to the Jeans mass) is expected to be 
as large as $\sim 100 \ \Msun$. It appears that the efficient cooling by
fine-structure lines at the low densities cannot be the main mechanism
to form Jeans-unstable fragments with small masses.
Hence, in order to produce low-mass stars in this scenario, protostars 
must be formed and ejected from such massive clumps by $N$-body
interactions before they grow to high masses \citep[e.g.,][]{Ji14}.
Although not entirely impossible, the scenario seems to require rather
specific conditions.
Intriguingly, other studies argue that the presence of dust grains is essential 
for the formation of the first low-mass stars; 
heat transfer via collisions from gas to dust effectively cools the gas and
induces the gravitational instability, 
to trigger the core fragmentation into small clumps.
This occurs at very high gas densities of $\nH \sim 10^{12}$--$10^{14} \ \percc$,
where the typical mass scale of fragments is in the range of $\sim 0.01$--$0.1 \ \Msun$
\citep[e.g.,][]{Schneider03, Schneider06}.
Assuming a certain grain size distribution and the dust-to-metal mass ratio
($\sim 0.5$) of the local interstellar medium, \citet{Omukai05} derive 
the critical metallicity of $Z_{\rm cr} \sim 10^{-5.5} \ \Zsun$,
above which the dust cooling can trigger the gas fragmentation.

It is important to note that the conditions for fragmentation of low-metallicity
gas clouds depend on the properties of dust grains, such as the size distribution
and the composition, which are expected to be different in the early universe.
For example, recent observations of damped Lyman $\alpha$ systems with metallicity 
$Z\sim 10^{-3} \ \Zsun$ reveal that the mass fraction of dust relative to metal is smaller 
than the present-day value \citep{Molaro00, DeCia13}.
In the local universe, dust grains are thought to be formed in
the stellar wind of asymptotic giant branch stars as well as in the ejecta of supernovae.
Subsequent accretion of heavy elements onto dust in molecular clouds also 
contributes the enhancement of dust mass fraction.
The prompt formation path of dust in the early universe is limited to the Pop III supernovae
whose progenitors have short life-times \citep{Todini01,Nozawa03}.
Newly formed grains in the supernova ejecta are, however, destroyed by sputtering 
after the reverse 
shocks penetrate into the ejecta \citep{Bianchi07, Nozawa07, Silvia10,Silvia12}.
Metals locked up in dust grains are partly returned into the gas phase.
Therefore, the mass ratio of grains to gas-phase metals can be significantly
small. It is thus important to study the role of dust grains in the first 
galaxies by employing realistic dust formation models.

\citet{Schneider12Crit, Schneider12Caf} investigate the fragmentation properties of gas clouds
pre-enriched with the grains that are produced by Pop III supernovae.
They show that the initial dust-to-gas mass ratio is a key quantity for the 
fragmentation condition.
The dust-induced fragmentation is not triggered if this ratio is significantly reduced
by the destruction of dust in the supernova ejecta.
The conclusion is drawn under the assumption that the size distribution and dust-to-gas 
mass ratio never change in the course of cloud collapse.
\citet{Nozawa12} point out that the growth of dust grains due to the accretion of heavy 
elements can take place and increase the dust-to-gas mass ratio even 
in very low-metallicity gas clouds.

\citet{Chiaki13Single} show that the cloud fragmentation can be triggered by the 
grain growth even if the initial fraction of heavy element condensed into grains is as small as 0.001.
When the grain growth is considered, 
the fragmentation conditions rely on the gas metallicity {\it and} the initial grain size
rather than only on the initial dust-to-gas mass ratio.
The estimated critical metallicity is 
$\Zcrit \sim 10^{-4.5}$ for the initial grain radii $r_0\geq 0.1 \ \um$,
and $\Zcrit \sim 10^{-5.5}$ for $r_0 \leq 0.01 \ \um$.
\citet{Chiaki13Single}, however, assume a single size and a single component of dust
($\Pyroxene$, enstatite)
which may be too simple as a realistic dust model in the early universe.
Clearly, further studies of the thermal evolution of the collapsing clouds
based on more realistic dust models are necessary to clarify the formation
condition of the first low-mass stars.

There are two leading studies that independently predict the size distribution and
the total amount of dust ejected from Pop III supernovae with a wide range of progenitor masses:
the models of \citet[hereafter \Noz]{Nozawa07} and \citet[hereafter \Sch]{Schneider12Crit}.
Both of the studies treat the formation of dust in the expanding ejecta and the destruction of dust
by the reverse shocks, but present different compositions, size distributions, and masses of dust.
Thus, the fragmentation condition can be different for the two dust models.
In this paper, we explore the thermal evolution of low-metallicity star forming clouds
by employing these two sets of supernova dust models and 
by taking into account the growth of multiple grain species with size distributions.

In section 2, we describe our one-zone model of the cloud evolution and 
the calculation of grain growth.
In Section 3, we discuss the effect of grain growth on the thermal evolution of clouds.
Then, we present the critical abundances of heavy elements for the formation of low-mass fragments.
Concluding remarks and discussion are given in Section 4.

%%%%%%%%%%%%%%%%%%%%%%%%%%%%%%%%%%%%%%%%%%%%%
% Numerical Simulations %%%%%%%%%%%%%%%%%%%%%%%%%%%%%%%%
%%%%%%%%%%%%%%%%%%%%%%%%%%%%%%%%%%%%%%%%%%%%%
\begin{table}
\caption{Silicon Chemical Reactions}
\label{tab:Si_chem}
\begin{tabular}{@{}rccc}
\hline
No. & Reaction & Rate Coef. & Ref. \\ 
\hline \hline
Si1 & Si + OH $\to$ SiO + H & $3\E{-11}$ & H80\\
Si2 & Si + O$_2$ $\to$ SiO + O & $1.3\E{-11} \exp (-111/T) $ & LG90 \\
Si3 & SiO + OH $\to$ SiO$_2$ + H & $2\E{-13}$ & H80 \\
\hline
\end{tabular}
\medskip
The units of temperature $T$ and rate coefficients are
K and ${\rm cm^3 \ s^{-1}}$, respectively.
Ref.: H80; \citet{Hartquist80}, LG90; \citet{Langer90}
\end{table}

\begin{figure}
\includegraphics[width=7cm]{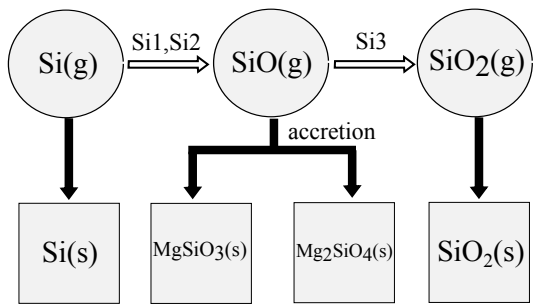}
\caption{
Chemical network of silicon (white arrows) which are included in our one-zone calculations.
We also consider accretion of gas-phase species onto grains (indicated by black arrows).
The reaction tabs above the arrows are identical with
the reaction numbers in Table \ref{tab:Si_chem}.
Circles and squares depict gas-phase (g) and solid-phase (s)
species, respectively.
}
\label{fig:Si_chem}
\end{figure}

\begin{figure*}
\includegraphics[width=18cm]{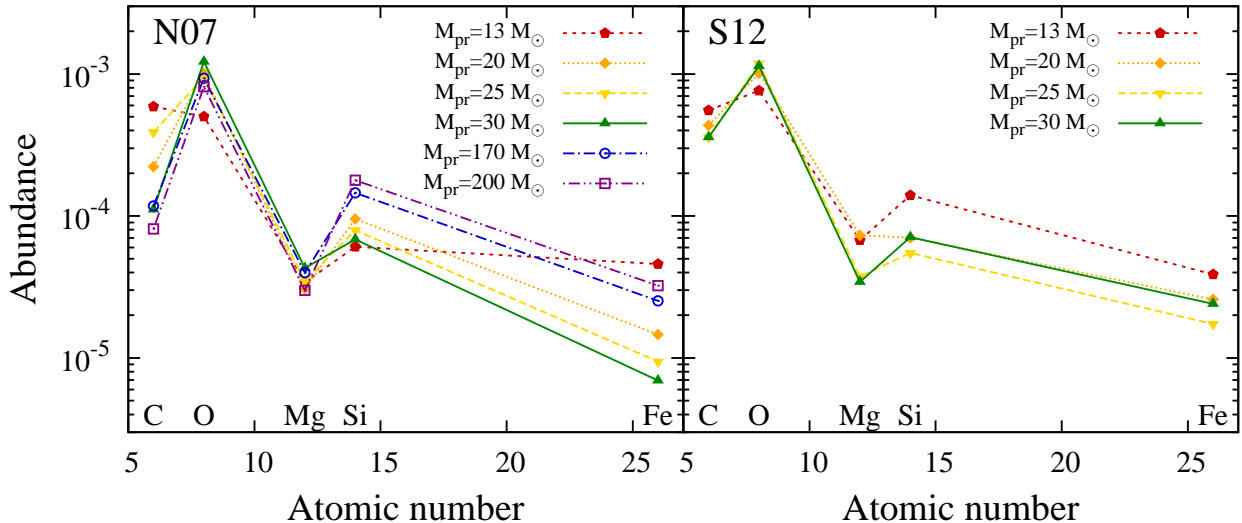}
\caption{
Number abundance $A_j$ of heavy element $j$ for the
Pop III supernova models of \Noz \ (left) and \Sch \ (right).
We plot the values derived by Equation (\ref{eq:Aj}) with $Z=Z_{\odot}$.
For each supernova model, red pentagons, orange diamonds, yellow opposite triangles, 
green triangles, blue circles, and purple squares show the abundances for supernovae with
progenitor masses 13, 20, 25, 30, 170, and 200 $\Msun$, respectively.
}
\label{fig:metal}
\end{figure*}

\begin{table*}
\caption{Grain Species Considered in the Calculations}
\label{tab:grains}
\begin{tabular}{lllrr}
\hline
Grains & Key species & Chemical Reaction & $\mu_{ij}$ & $a_{ij,0}$ (\AA) \\
\hline \hline
Si(s) & Si(g) & Si(g)$\to$ Si(s) & 28.0 & 1.684 \\
Fe(s) & Fe(g) & Fe(g)$\to$ Fe(s) & 56.0 & 1.411 \\
$\Olivine$(s) & Mg(g) &  2Mg(g) + SiO(g) + 3H$_2$O(g) $\to$ $\Olivine$(s) + 3H$_2$(g) & 70.0 & 2.055 \\
& SiO(g) &  2Mg(g) + SiO(g) + 3H$_2$O(g) $\to$ $\Olivine$(s) + 3H$_2$(g) & 140.0 & 2.589 \\
$\Pyroxene$(s) & Mg(g), SiO(g) &  Mg(g) + SiO(g) + 2H$_2$O(g) $\to$ $\Pyroxene$(s) + 2H$_2$(g) & 100.0 & 2.319 \\
$\Magnetite$(s) & Fe(g) & 3Fe(g) + 4H$_2$O(g) $\to$ $\Magnetite$(s) + 4H$_2$(g) & 77.3 & 1.805 \\
C(s) & C(g) &C(g)$\to$ C(s)  & 12.0 & 1.281 \\
$\Silica$(s) & $\Silica$(g) & $\Silica$(g)$\to$ $\Silica$(s) & 60.0 & 2.080 \\
$\Magnesia$(s) & Mg(g) & Mg(g) + H$_2$O(g) $\to$ $\Magnesia$(s) + H$_2$(g) & 40.0 & 1.646 \\
$\Troilite$(s) & Fe(g), S(g) & Fe(g) + S(g) $\to$ $\Troilite$(s) & 88.0 & 1.932 \\
$\Alumina$(s) & Al(g) & 2Al(g) + 3H$_2$O(g) $\to$ $\Alumina$(s) + 3H$_2$(g) & 51.0 & 1.718 \\
\hline
\end{tabular}
\\ \medskip
The subscripts ``(s)'' and ``(g)'' are attached to solid- and gas-phase species, respectively.
The values are taken from \citet{Nozawa03}. $\mu _{ij}$ and $a_{ij,0}$ are 
the molecular weight and the hypothetical radius of a monomer molecule of grain species $i$
per nuclei of the key element $j$. 
\end{table*}

\begin{figure*}
\includegraphics[width=16cm]{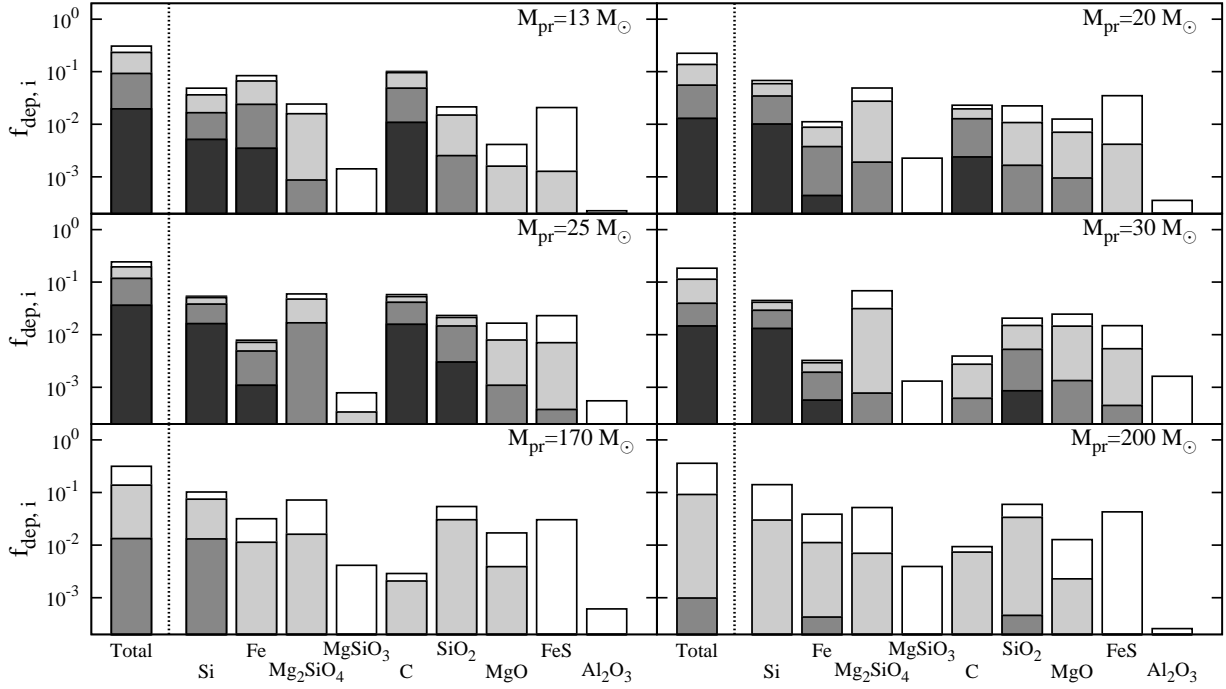}
\caption{
Mass fraction $f_{{\rm dep},i}$ of grain species $i$
relative to the total metal mass for supernova model of \Noz.
Each histogram represents the mass fraction for n0, n0.1, n1, 
and n10 model from top to bottom.
The first histogram represents the depletion factor $\fdep $
whereas the other histograms show the contribution of each dust species.
}
\label{fig:dust_Nu}
\end{figure*}

\begin{figure*}
\includegraphics[width=16cm]{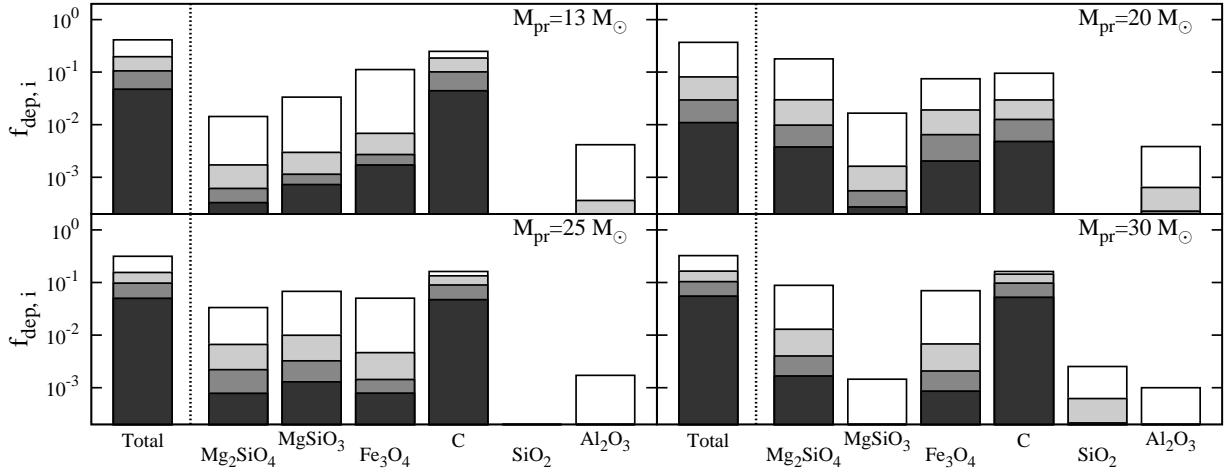}
\caption{
Same as Figure \ref{fig:dust_Nu} but for models by \Sch.
}
\label{fig:dust_S0}
\end{figure*}

\section{NUMERICAL METHOD}

\subsection{Collapse model and gas-phase chemistry} 

We follow the evolution of the cloud temperature $T$ along with the
increasing density calculated by a one-zone semi-analytic collapse model of \citet{Omukai00}.
In this paper, the cloud density is described as the 
hydrogen number density $\nH = \XH \rho / \mH$, where 
$\XH$ is the hydrogen mass fraction,
$\rho$ is the total (gas and dust) mass density, and
$\mH$ is the mass of a hydrogen atom.
We include in this model non-equilibrium gas chemistry of eight 
species of primordial elements
H$^+$, $e^-$, H, H$^-$, H$_2$, D$^+$, D, and HD,
and nineteen species made of heavy elements
C$^+$, C, CH, CH$_2$, CO$^+$, CO, CO$_2$, O$^+$, O, OH$^+$, OH, 
H$_2$O$^+$, H$_2$O, H$_3$O$^+$, O$_2^+$, O$_2$, Si, SiO, and SiO$_2$.
We solve the reduced chemical networks of \citet{Omukai10}
supplemented with the silicon chemistry.
For species containing silicon, oxidation of Si and SiO is considered as
the major reactions. 
Table \ref{tab:Si_chem} and Figure \ref{fig:Si_chem} present the gas-phase chemical reactions of silicon and
the rate coefficients used in this work.
We also solve the cooling rates by emission lines of atoms/ions (C {\sc i}, C {\sc ii}, and
O {\sc i}), and by molecules (H$_2$, HD, CO, OH, and H$_2$O).
When the cloud becomes optically thick, the cooling rate from each line is reduced
by the corresponding escape fraction.
To this end, we determine the column density assuming the size of the cloud to be 
given by the Jeans length.

We set the initial hydrogen number density and temperature
$n_{{\rm H},0}=0.1 \ \percc$ and $T_0= 300 \ {\rm K}$, respectively.\footnote{Hereafter, the subscript `0' is 
referred to as the initial values. }
The initial number abundances of H$^+$ and H$_2$ relative to hydrogen nuclei are
$y_0 ({\rm H^+})= 10^{-4}$ and $y_0 ({\rm H_2})= 10^{-6}$.
The total number abundances of deuterium and helium are 
$3.0\E{-5}$ and $0.083$
(corresponding to the mass fraction $Y_{{\rm He}}= 0.25$), respectively.
We assume that elements C, O, and Si are
initially in the form of C$^+$ ions, neutral O atoms and Si atoms, respectively.

We consider the clouds to be enriched by Pop III supernovae.
The supernova models give the mass yield $M_j$ of heavy element $j$ and
the total metal yield $M_{{\rm metal}} = \sum M_j$.
Pop III supernovae are generally characterized by metallicities in the range 1--30 $\Zsun$
\citep{Umeda02,Limongi12}.
Heavy elements are diluted within the expanding ejecta and eventually mixed with the ambient
primordial gas.
Thus, the formation site of the second-generation stars, polluted with the metal and dust has
typically a lower metallicity than the solar value.
Setting the cloud metallicity $Z$ as an effective dilution factor,
we obtain the number abundance of heavy element $j$ relative to hydrogen nuclei in the cloud as
\begin{equation}
A_j = \frac{Z}{\mu _j \XH } \frac{M_{j}}{M_{{\rm metal}}},
\label{eq:Aj}
\end{equation}
where $\mu _j$ is the molecular weight of element $j$,
and $\XH = 1-Y_{\rm He}$ is the hydrogen mass fraction.

Both \Noz \ and \Sch \ employ the core-collapse supernova (CCSN) models with progenitor masses
$\Mpr = 13$, 20, 25, and $30 \ \Msun$ (hereafter called M13, M20, M25, and M30 models,
respectively).
\Noz \ further explore the pair-instability supernova (PISN) models
with $\Mpr = 170$ (M170) and $200 \ \Msun$ (M200).
%\Sch \ adjust the parameters such as the explosion energy, the mass cut, and the $\rm ^{56} Ni$
%decay rate to reconstruct the abundance pattern of
%the most primitive star \Caf \ \citep{Caffau11Nat}.
\Sch \ adjust the mass cut, for each SN model, in order to fit the abundance pattern of the most 
primitive star \Caf \ \citep{Caffau11Nat}.
Such a procedure, as well as the properties of each progenitor mass and explosion details, 
has been extensively explained  in \citet{Limongi12}.
Figure \ref{fig:metal} shows the abundances of the major heavy elements for various
progenitor masses.
For all the progenitor models, the abundances deviate from the solar values by $\sim 0.5$ dex.
While \Sch \ model can predict the abundances less sensitive to the progenitor mass,
\Noz \ model presents the various patterns.
N07M13 models predict $\rm C>O$, and heavier elements such as Si and Fe increases
with the increasing progenitor mass.

\subsection{Dust models}

\subsubsection{Dust properties}

We consider ten dust species:
silicon (Si), iron (Fe), forsterite ($\Olivine$), enstatite ($\Pyroxene$), magnetite ($\Magnetite$),
amorphous carbon (C),
silica ($\Silica$), magnesia ($\Magnesia$), troilite ($\Troilite$), and alumina ($\Alumina$).
Note that, depending on the employed supernova models,
some of the species are not efficiently formed in the supernova ejecta, or are fully destroyed by the reverse shocks.

We quantify the amount of dust grains by the condensation efficiency $f_{ij}$, 
which is defined as the number fraction
of nuclei of element $j$ locked in dust species $i$.\footnote{
In this paper, the subscripts $i$ and $j$ denote grain species and heavy elements, respectively.}
By using the condensation efficiency, the mass density of dust species $i$ in the cloud is written as 
$
\rho_i = f_{ij} A_j \nH \mu _{ij} \mH,
$
where $\mu_{ij}$ is the molecular weight of the grain species $i$
per nucleus of element $j$ (see Table \ref{tab:grains}).

In this study, we introduce the differential size distribution function $\varphi _i (r)$
of grain species $i$ normalized as $\int \varphi _i(r) dr = 1$.
Then, the number of dust particles per unit volume is
\begin{equation}
n_i = \frac{\rho_i}{(4\pi /3) s_i \int r^3 \varphi _i(r) dr},
\label{eq:n_dust}
\end{equation}
where $s_i$ is the bulk density of an individual dust particle
derived from the values of $\mu_{ij}$ and $a_{ij,0}$ in Table \ref{tab:grains}.
We can calculate the number density of grains with radii between $r$ and $r+dr$ 
as $n_i \varphi_i(r) dr$.

\subsubsection{H$_2$ formation on grains}

In the presence of dust grains,
H$_2$ molecules are efficiently formed on grain surfaces.
Since H$_2$ molecules are major coolants in a low-metallicity gas,
the thermal evolution of the collapsing gas is significantly affected (especially in the early stages of collapse)
by the amount of dust.
The formation rate of hydrogen molecules per grain of species $i$ with radius $r$ is
\begin{equation}
{\cal R}_{{\rm H_2},i}(r) = \frac 12 n({\rm H_I}) \langle v_{{\rm H}} \rangle 
\pi r^2 \epsilon _{{\rm H}_2} S_{{\rm H}} ,
\label{DustEvaporation}
\end{equation}
where $n(x) $ is the number density of gas-phase chemical species $x$,
$\langle v_{{\rm H}} \rangle = (8kT/\pi m_{{\rm H}}) ^{1/2}$ is the average 
velocity of hydrogen atoms,
$\epsilon _{{\rm H}_2}$ is the efficiency of H$_2$ recombination on grain surfaces,
and $S_{{\rm H}}$ is the sticking efficiency of 
hydrogen atoms impacting grain surfaces
\citep{CazauxTielens02}. 
The values of $\epsilon _{{\rm H}_2}$
and $S_{{\rm H}}$ are 
functions of both gas and grain temperatures
\citep[see][for detailed formulation]{Schneider06}.
We calculate ${\cal R}_{{\rm H_2},i} (r)$ for each dust species and each grain radius.
Then, the formation rate of hydrogen molecules on grain
surfaces per unit volume is described as
\begin{equation}
\left. \frac{dn_{{\rm H_2}}}{dt} \right|_{{\rm on \ grains}}=
\sum _i \int {\cal R}_{{\rm H_2},i}(r) n_i \varphi _i(r) dr.
\end{equation}

\subsubsection{Dust temperature and cooling}

The temperature $T_i(r)$ of grain species $i$ with radius $r$ can be derived
from the balance between the heating by collisions with the gas particles
(mostly hydrogen and helium atoms and electrons) and 
the cooling by thermal emission of dust.
We ignore the heat exchange among dust particles 
because the grain-grain collision rate
is much smaller than
the gas-grain collision rate in the low metallicity environments considered in this paper.
The heating rate of a dust grain with radius $r$ owing to collisions with the gas particles is
\begin{equation}
{\cal G} _i (r) = \pi r^2 
\langle n v_{{\rm g}} \rangle  [2kT-2kT_i(r)],
\label{dust:Gamma}
\end{equation}
where $\langle n v_{{\rm g}}\rangle =
[n({\rm H_I}) + n({\rm H_2})/\sqrt{2} + n({\rm He_I})/2]
(8kT/\pi m_{{\rm H}})^{1/2}$ 
is the average velocity of gas particles.
The cooling rate of a grain through thermal radiation is
\begin{equation}
{\cal L} _i (r)
=  4 \sigma _{{\rm B}} T_i^4(r) \left\langle Q^{\nu}_i(r) \right\rangle \pi r^2 \beta _{{\rm cont}},
\label{dust:Lambda}
\end{equation}
where $\sigma _{{\rm B}}$ is the Stephan-Boltzmann coefficient,
$\left\langle Q^{\nu}_i(r) \right\rangle $ is the Planck-mean of the absorption coefficient,
and $\beta _{{\rm cont}}$ is the continuum escape fraction 
(see the next Section).
From the energy balance equation of a dust grain, 
${\cal G} _i(r) = {\cal L} _i(r)$,
we can obtain dust temperature for each dust species and size.
Then, the cooling rate of the gas owing to dust per unit volume
is calculated by summing up Equation (\ref{dust:Lambda}) over
all sizes and species:
\begin{equation}
\Lambda _{{\rm d}}
 = \sum _i \int {\cal L}_i (r) n_i \varphi _i(r) dr.
\label{dust:totLambda}
\end{equation}

\subsubsection{Continuum opacity}

The Planck-mean absorption coefficients of the grain species are taken 
from the references in Table 1 of \citet{Nozawa08}.
The continuum optical depth is
\begin{equation}
\tau _{{\rm cont}}= \left( \kappa _{{\rm g}} \rho _{{\rm g}} + 
\sum _{i} \int \left\langle Q^{\nu}_i(r) \right\rangle \pi r^2 n_i \varphi _i(r) dr \right) l_{{\rm sh}},
\end{equation}
where $\kappa _{{\rm g}}$ is the Planck opacity of gas taken from \citet{Mayer05},
and $\rho _{{\rm g}} = \rho - \sum _i \rho _i$ is the mass density of gas-phase species.
We calculate the shielding length $l_{{\rm sh}}$ as in \citet{Omukai00}.
Then, the escape fraction of continuum emission is calculated as
$\beta _{{\rm cont}}=\min \{ 1, \tau _{{\rm cont}}^{-2} \}$ \citep{Omukai00}.

\subsubsection{Grain growth}

We consider that the grain growth by the accretion of the gas species
proceeds via the reactions in Table \ref{tab:grains}.
Here, we assume that the heavy element species with the smallest timescale
of collisions with grains controls the kinetics of grain growth 
(hereafter, referred as {\it key species}).
In this model, silicon, silicate (forsterite and enstatite), and
silica grains grow through the accretion of Si atoms, SiO molecules,
and SiO$_2$ molecules, respectively,
as indicated by filled arrows in Figure \ref{fig:Si_chem}.
The flow of silicon from the gas phase to the solid phase is shown in Figure
\ref{fig:Si_chem}, along with the gas phase reactions.

The growth rate of grain radius is given by
\begin{equation}
\left( \frac{dr}{dt} \right) _i = \alpha _i \left( \frac{4\pi}{3} a_{ij,0}^3 \right)
\left( \frac{kT}{2\pi m_{i1}} \right)^{1/2} n_{i1} (t),
\label{GG:drdt} 
\end{equation}
where $\alpha _i$ denotes the sticking probability of the gaseous species
incident onto surfaces of grains $i$, and
$a_{ij,0}$ denotes the hypothetical radius of a monomer molecule of grain species $i$ 
in the dust phase, and
$m_{i1}$ and $n_{i1} $ respectively denote the mass and the number density of the key species $i1$. 
We here assume the sticking probability $\alpha _i = 1$
(see Section 4 for the justification of the select of this value and for the effect
of smaller $\alpha _i$ on the results).
Note that the growth rate does not depend on the grain radius $r$.
Therefore, the size distribution $\varphi _i(r,t)$ at a time $t$ is equivalent to
the initial one $\varphi_{i,0}(r)$ shifted to larger radii by 
$\Delta r = \int _{0}^{t} (dr/dt)_i dt$.
Then, the time evolution of the condensation efficiency is given as
\begin{equation}
f_{ij}(t) = f_{ij,0} 
\frac{\int  r^3 \varphi _i (r,t) dr}{\int r^3 \varphi _{i,0} (r) dr}.
\label{GG:f}
\end{equation}
The density $n(x_j)$ of the gas-phase species $x_j$ containing the element $j$ 
is determined by the condensation of the element $j$ onto all the relevant grain species $i$ as
$n(x_j, t+\Delta t)= n(x_j,t) -\sum_i \left\{ f_{ij}(t+\Delta t) - f_{ij}(t) \right\}  A_j n_{\rm H}$ 
as well as the relevant gas-phase chemistry.

\begin{figure}
\includegraphics[width=9cm]{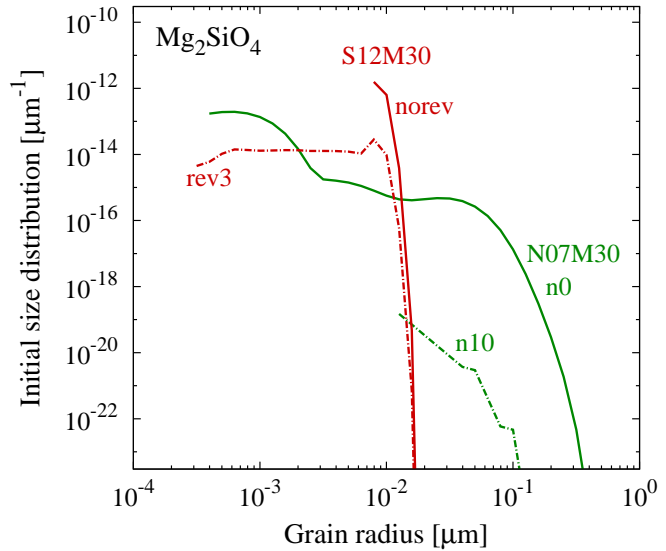}
\caption{
Initial size distribution of $\Olivine$ grains for the \Noz \ (green) and \Sch \ (red)
models without the effect of dust destruction (solid) and with the largest efficiency
of dust destruction among their reverse shock models (dot-dashed).
}
\label{fig:size}
\end{figure}

\begin{table*}
\caption{Initial Values for Supernova Models and the Critical Conditions without Grain Growth
\label{tab:Init_Val}}
\begin{tabular}{@{}lclrrrrrrrr}
\hline
Model & $\Mpr$ & rev. &
$f_{\rm dep,C,0}$ &
$S_{\rm C,0}$ &
$f_{\rm dep,Sil,0}$ &
$S_{\rm Sil,0}$ &
$f_{\rm dep,0}$ &
${\cal D}_0$ & 
${\cal D}_{\rm cr,ng}$ & 
$[Z_{\rm cr,ng}]$ \\
\hline 
N07M13n0    & 13 &   n0 & ${\bf  0.101}$ & ${\bf  2.65}$ & $      0.024 $ & $      3.83 $ & $ 0.305$ & $ 6.106$ & $ 3.060$ & $  -5.3$ \\
N07M13n0.1  &    & n0.1 & ${\bf  0.096}$ & ${\bf  2.64}$ & $      0.016 $ & $      2.77 $ & $ 0.233$ & $ 4.659$ & $ 2.335$ & $  -5.3$ \\
N07M13n1    &    &   n1 & ${\bf  0.049}$ & ${\bf  2.50}$ & $     <0.001 $ & $      3.76 $ & $ 0.093$ & $ 1.851$ & $ 2.330$ & $  -4.9$ \\
N07M13n10   &    &  n10 & ${\bf  0.011}$ & ${\bf  3.88}$ & $     <0.001 $ & $      7.77 $ & $ 0.020$ & $ 0.392$ & $ 1.966$ & $  -4.3$ \\
N07M20n0    & 20 &   n0 & $      0.023 $ & $      4.51 $ & ${\bf  0.049}$ & ${\bf  4.22}$ & $ 0.220$ & $ 4.402$ & $ 2.777$ & $  -5.2$ \\
N07M20n0.1  &    & n0.1 & ${\bf  0.020}$ & ${\bf  3.46}$ & ${\bf  0.028}$ & ${\bf  2.96}$ & $ 0.137$ & $ 2.750$ & $ 3.462$ & $  -4.9$ \\
N07M20n1    &    &   n1 & ${\bf  0.013}$ & ${\bf  1.96}$ & $      0.002 $ & $      3.61 $ & $ 0.055$ & $ 1.110$ & $ 5.563$ & $  -4.3$ \\
N07M20n10   &    &  n10 & ${\bf  0.002}$ & ${\bf  2.24}$ & $     <0.001 $ & $      4.87 $ & $ 0.013$ & $ 0.259$ & $ 8.199$ & $  -3.5$ \\
N07M2513n0  & 25 &   n0 & ${\bf  0.058}$ & ${\bf  2.42}$ & ${\bf  0.060}$ & ${\bf  3.53}$ & $ 0.240$ & $ 4.797$ & $ 3.810$ & $  -5.1$ \\
N07M25n0.1  &    & n0.1 & ${\bf  0.053}$ & ${\bf  1.91}$ & ${\bf  0.048}$ & ${\bf  1.42}$ & $ 0.195$ & $ 3.907$ & $ 3.907$ & $  -5.0$ \\
N07M25n1    &    &   n1 & ${\bf  0.041}$ & ${\bf  1.12}$ & $      0.017 $ & $      1.35 $ & $ 0.118$ & $ 2.351$ & $ 4.692$ & $  -4.7$ \\
N07M25n10   &    &  n10 & ${\bf  0.016}$ & ${\bf  0.83}$ & $     <0.001 $ & $      3.05 $ & $ 0.036$ & $ 0.726$ & $ 4.578$ & $  -4.2$ \\
N07M30n0    & 30 &   n0 & $      0.004 $ & $      6.85 $ & ${\bf  0.068}$ & ${\bf  3.99}$ & $ 0.184$ & $ 3.671$ & $ 2.916$ & $  -5.1$ \\
N07M30n0.1  &    & n0.1 & $      0.003 $ & $      7.14 $ & ${\bf  0.031}$ & ${\bf  4.04}$ & $ 0.113$ & $ 2.265$ & $ 3.590$ & $  -4.8$ \\
N07M30n1    &    &   n1 & $     <0.001 $ & $     10.27 $ & $     <0.001 $ & $      8.57 $ & $ 0.040$ & $ 0.791$ & $ 6.280$ & $  -4.1$ \\
N07M30n10   &    &  n10 & $     <0.001 $ & $      7.52 $ & $     <0.001 $ & $      7.17 $ & $ 0.015$ & $ 0.292$ & $11.639$ & $  -3.4$ \\
N07M170n0   &170 &   n0 & $      0.003 $ & $      4.66 $ & ${\bf  0.072}$ & ${\bf  6.62}$ & $ 0.315$ & $ 6.297$ & $ 3.156$ & $  -5.3$ \\
N07M170n0.1 &    & n0.1 & $      0.002 $ & $      4.88 $ & $      0.016 $ & $      4.76 $ & $ 0.138$ & $ 2.761$ & $ 4.377$ & $  -4.8$ \\
N07M170n1   &    &   n1 & $     <0.001 $ & $     10.03 $ & $     <0.001 $ & $      1.27 $ & $ 0.013$ & $ 0.266$ & $16.796$ & $  -3.2$ \\
N07M170n10  &    &  n10 & $     <0.001 $ & $     12.31 $ & $      0.000 $ &           --- & $<0.001$ & $ 0.001$ & $ 1.488$ & $ >-2.0$ \\
N07M200n0   &200 &   n0 & $      0.009 $ & $      5.39 $ & ${\bf  0.052}$ & ${\bf  9.29}$ & $ 0.359$ & $ 7.189$ & $ 3.603$ & $  -5.3$ \\
N07M200n0.1 &    & n0.1 & $      0.007 $ & $      5.20 $ & $      0.007 $ & $      4.94 $ & $ 0.091$ & $ 1.824$ & $ 2.892$ & $  -4.8$ \\
N07M200n1   &    &   n1 & ${\bf <0.001}$ & ${\bf 15.37}$ & $     <0.001 $ & $      1.60 $ & $<0.001$ & $ 0.020$ & $ 4.935$ & $  -2.6$ \\
N07M200n10  &    &  n10 & $     <0.001 $ & $     18.02 $ & $      0.000 $ &           --- & $<0.001$ & $<0.001$ & $<0.001$ & $ >-2.0$ \\
\hline                                                                                  
S12M13norev & 13 & norev& $      0.249 $ & $      4.59 $ & $      0.033 $ & $     45.76 $ & $ 0.413$ & $ 8.260$ & $ 0.826$ & $  -6.0$ \\
S12M13rev1  &    &  rev1& ${\bf  0.185}$ & ${\bf  3.35}$ & $      0.003 $ & $     53.73 $ & $ 0.197$ & $ 3.947$ & $ 1.571$ & $  -5.4$ \\
S12M13rev2  &    &  rev2& ${\bf  0.101}$ & ${\bf  3.22}$ & $      0.001 $ & $     52.36 $ & $ 0.106$ & $ 2.117$ & $ 1.336$ & $  -5.2$ \\
S12M13rev3  &    &  rev3& ${\bf  0.044}$ & ${\bf  3.28}$ & $     <0.001 $ & $     47.14 $ & $ 0.047$ & $ 0.943$ & $ 1.494$ & $  -4.8$ \\
S12M20norev & 20 & norev& ${\bf  0.095}$ & ${\bf 37.56}$ & ${\bf  0.179}$ & ${\bf 14.68}$ & $ 0.369$ & $ 7.384$ & $ 0.738$ & $  -6.0$ \\
S12M20rev1  &    &  rev1& ${\bf  0.030}$ & ${\bf 32.77}$ & ${\bf  0.030}$ & ${\bf 18.12}$ & $ 0.081$ & $ 1.618$ & $ 0.811$ & $  -5.3$ \\
S12M20rev2  &    &  rev2& ${\bf  0.013}$ & ${\bf 27.85}$ & $      0.010 $ & $     18.22 $ & $ 0.030$ & $ 0.591$ & $ 0.937$ & $  -4.8$ \\
S12M20rev3  &    &  rev3& ${\bf  0.005}$ & ${\bf 23.72}$ & $      0.004 $ & $     17.00 $ & $ 0.011$ & $ 0.219$ & $ 1.098$ & $  -4.3$ \\
S12M25norev & 25 & norev& ${\bf  0.162}$ & ${\bf 11.51}$ & ${\bf  0.068}$ & ${\bf 29.42}$ & $ 0.315$ & $ 6.293$ & $ 0.997$ & $  -5.8$ \\
S12M25rev1  &    &  rev1& ${\bf  0.134}$ & ${\bf  3.49}$ & ${\bf  0.010}$ & ${\bf 34.48}$ & $ 0.155$ & $ 3.106$ & $ 2.467$ & $  -5.1$ \\
S12M25rev2  &    &  rev2& ${\bf  0.090}$ & ${\bf  2.32}$ & ${\bf  0.003}$ & ${\bf 34.39}$ & $ 0.097$ & $ 1.934$ & $ 2.435$ & $  -4.9$ \\
S12M25rev3  &    &  rev3& ${\bf  0.047}$ & ${\bf  2.04}$ & $      0.001 $ & $     33.17 $ & $ 0.050$ & $ 1.000$ & $ 2.512$ & $  -4.6$ \\
S12M30norev & 30 & norev& $      0.161 $ & $      1.44 $ & ${\bf  0.088}$ & ${\bf 26.22}$ & $ 0.324$ & $ 6.485$ & $ 1.629$ & $  -5.6$ \\
S12M30rev1  &    &  rev1& ${\bf  0.144}$ & ${\bf  1.12}$ & ${\bf  0.013}$ & ${\bf 31.72}$ & $ 0.164$ & $ 3.285$ & $ 2.609$ & $  -5.1$ \\
S12M30rev2  &    &  rev2& ${\bf  0.097}$ & ${\bf  1.15}$ & ${\bf  0.004}$ & ${\bf 32.36}$ & $ 0.104$ & $ 2.071$ & $ 2.607$ & $  -4.9$ \\
S12M30rev3  &    &  rev3& ${\bf  0.052}$ & ${\bf  1.18}$ & ${\bf  0.002}$ & ${\bf 28.84}$ & $ 0.055$ & $ 1.094$ & $ 2.182$ & $  -4.7$ \\
\hline
\end{tabular}
\\ \medskip
Note --- 
Progenitor mass $\Mpr$ is in the unit of $\Msun$.
The third column `rev.' refers to the reverse shock model (see text),
$f_{{\rm dep},i,0}$ is the initial mass fraction of grain species $i$ relative to the total metal mass.
and $S_{i,0}$ is the initial geometrical cross-section of grain $i$ per unit
dust mass ($\E{4} \ {\rm cm^2 \ g^{-1}}$).
We write $f_{{\rm dep},i,0}$ and $S_{i,0}$ in bold if either carbon or silicate dust is the dominant coolant
for each model.
For silicate grains ($i={\rm Sil}$), we show the values for the dominant species: 
$\Pyroxene$ for S12M13 and S12M25 models, and $\Olivine$ for the other models.
${\cal D}_0$ is the dust-to-gas mass ratio for our $Z=10^{-5} \ \Zsun$ calculations ($\E{-8}$).
${\cal D}_{\rm cr,ng}$ and $Z_{\rm cr,ng}$
are the critical dust-to-gas mass ratio and metallicity
which we determine by one-zone calculations without grain growth
 ($[Z_{\rm cr,ng}] = \log (Z_{\rm cr,ng} / \Zsun)$).
Since we set metallicities varying every 0.1 dex, 
the value $[Z_{\rm cr,ng}]=-5.9$ indicates, for example, 
that fragmentation condition is met at $[Z_{\rm cr,ng}]\geq -5.9$, but not for $[Z_{\rm cr,ng}]\leq -6.0$.
\end{table*}

\subsection{Supernova dust models}

\Noz \ and \Sch \ calculate the formation of grains in the supernova ejecta for different progenitor masses,
predicting the composition and size distribution of newly-formed grains,
hereafter called n0 and norev models for \Noz \ and \Sch , respectively. 
For each dust formation model, they also study dust destruction by reverse shocks 
which occurs after the grain formation.
The strength of the reverse shock is parametrized with the density
of the ambient gas around the supernovae.
\Noz \ investigate the reverse shock models when the number densities of the ambient gas is 
$\namb = 0.1$, 1, and $10 \ \percc$. 
We call these models as n0.1, n1, and n10, respectively.
\Sch \ investigate for the mass densities of the ambient gas 
$\rhoamb = 10^{-25}$ (rev1), $10^{-24}$ (rev2), and $10^{-23} \ {\rm g \ cm^{-3}}$ (rev3).

These models give the mass $M_{i}$ of grain species $i$ surviving
against the destruction by the reverse shocks.
Figures \ref{fig:dust_Nu} and \ref{fig:dust_S0} show the ratio $f_{{\rm dep},i}$ 
of dust mass to the total metal mass for \Noz \ and \Sch \ models, respectively.
Table \ref{tab:Init_Val} shows the values for carbon and silicate.
One can see that the dust mass decreases with the increasing ambient gas density.
We can also calculate the initial condensation 
efficiency of the collapsing cloud as 
\begin{equation}
f_{ij,0} = \frac{M_{i}/\mu _{ij}}{M_{j}/\mu _j}.
\end{equation}

The efficiency of gas cooling is determined by the dust composition and 
size distribution.
\Noz \ and \Sch \ obtain different dust compositions.
First, the mass fraction of carbon grains is generally larger for \Sch \ model than \Noz \ model.
Carbon grains are formed by condensation of carbon atoms that are not oxidized to form CO
molecules in the supernova ejecta. 
\Sch \ consider the molecular destruction by the collision with high energy 
electrons from $^{56}$Co \citep{Todini01}.
\Noz \ do not include the dissociation of the molecules.
Although carbon grains are formed in the outer layers with $\rm C>O$ in the ejecta,
most of the carbon nuclei are in the inner layers, where oxygen nuclei are also abundant.
Thus, the formation of carbon grains is mitigated in \Noz \ model.

Second, magnetite grains are produced only in \Sch \ models.
This results from the different ejecta models.
Our \Noz \ model is taken from their unmixed ejecta model, where
the ejecta is considered to remain stratified, i.e., keeping the original onion-like structure
of composition of heavy elements during the explosions.
While the oxygen-rich layer is originally in the outer region, Fe is
in the innermost layer.
In this case, Fe$_3$O$_4$ grains are not formed because of the assumed inefficient mixing.
On the other hand, \Sch \ assume the fully mixed ejecta, where the composition of heavy 
elements is uniform, and
magnetite grains can be formed.

\Noz \ and \Sch \ models also produce different size distribution functions.
In Figure \ref{fig:size}, 
we compare the initial size distributions of $\Olivine $ grains for the two extreme cases:
without dust destruction (n0 and norev) 
and with the largest efficiency of the dust destruction among their reverse shock models (n10 and rev3).
%Given a thermal evolution and an elemental composition in the formation site of grains,
%the newly formed grains have a log-normal size distribution \citep{Nozawa03}.
%Under their assumption of the uniform elemental composition in the ejecta,
%\Sch \ predict approximately one single log-normal distribution for each grain species
%with a peak at the characteristic radius (see red solid curve in Figure \ref{fig:size}).
%The dust size for \Noz \ model is described by a superposition of log-normal functions.
%In their unmixed ejecta, a grain species is formed in the different layers with the various 
%temperature evolutions and elemental compositions.
%Therefore, the size distribution spreads in a wider range of radii than for \Sch \ model
%(green solid curve).
The size distribution of newly formed grains is mainly regulated by the 
number density of condensible gas species, i.e., elemental composition at the formation 
site of the grains.
For the dust model by \Sch, where the uniform elemental composition in the ejecta is assumed, 
the size distribution of newly formed grains is confined to a narrow range of grain radius (see red 
solid curve in Figure 5).
On the other hand, in the unmixed ejecta applied by \Noz, dust grains form with different 
characteristic radii in the layers with the different elemental compositions.
Thus, the resulting size distribution, which is made of the contributions of grains formed 
in each formation region, spreads in a wider range of radius than \Sch \ model (green solid curve).
%Given a thermal evolution and an elemental composition in the formation site of grains,
%the newly formed grains have an size distribution with a peak at the radius characterized
%by the gas temperature and elemental composition.
%\Sch \ predict such a distribution with a peak (see red solid curve in Figure \ref{fig:size}) from
%the uniformly mixed ejecta.
%On the other hand, in the unmixed ejecta in \Noz \ model, a grain species is formed in the different
%layers with the various temperature and elemental compositions,
%and thus the size distribution spreads in a wider range of radii than \Sch \ model (green solid curve).

The successive process of dust destruction also affects the size distribution of surviving grains.
\Sch \ assume that the grains are trapped in the high-temperature region defined by the reverse 
and forward shocks, where the grains undergo sputtering by the impact of high energy ions.
The grain size continues to decrease until the ejecta cools down sufficiently, and
the size distribution of surviving dust has a flat tail at the smaller radii
(see red dot-dashed curve in Figure \ref{fig:size}).
\Noz \ consider the relative motion between the expanding gas and dust grains.
The erupted dust grains are decelerated by the drag force from the gas.
Smaller grains are more tightly coupled with the gas so that
they remain in the shocked hot ejecta, continuing to be destroyed.
Some of the grain species are totally destroyed and returned to the gas phase.
Because larger grains are more likely to escape from the ejecta because of their larger inertia,
they survive without being significantly destroyed.
As a result, the size distribution is truncated at a given radius (green dot-dashed curve).

These different dust properties can affect the 
dust amount above which dust cooling activates the gas fragmentation:
for \Sch \ model, which predicts the smaller grain radii than \Noz, 
the efficiency of gas cooling by grains is expected
to be larger because the total grain cross-section is larger with the fixed dust mass.
In this paper, we define the critical condition for \Noz \ and \Sch \ supernova dust models
separately.
We consider the dust models
where grains are not destroyed by reverse shocks (n0 and norev), and also three
dust models where the reverse shock destruction is calculated for different ambient gas densities
for each progenitor mass.
In order to see the effect of grain growth, we study both cases with and without grain 
growth for each supernova dust model.
We determine the critical dust amount by varying the gas metallicity
in the range of $Z=10^{-7}$--$10^{-2} \ \Zsun$, where $Z_{\bigodot} = 0.02$.

%%%%%%%%%%%%%%%%%%%%%%%%%%%%%%%%%%%%%%%%%%%%%
% RESULTS %%%%%%%%%%%%%%%%%%%%%%%%%%%%%%%%%%%%%%
%%%%%%%%%%%%%%%%%%%%%%%%%%%%%%%%%%%%%%%%%%%%%

\begin{figure}
\includegraphics[width=9cm]{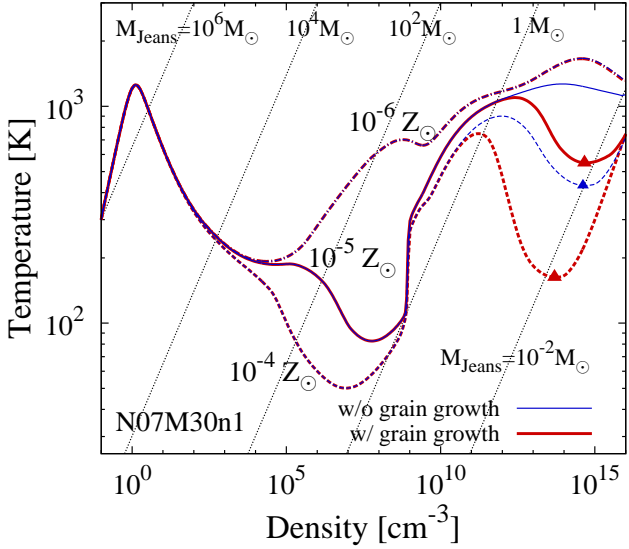}
\caption{
Temperature evolution as a function of the hydrogen number density
of the cloud centers for our N07M30n1 model with
metallicities $Z=10^{-6} \ \Zsun$ (dot-dashed),
$10^{-5} \ \Zsun$ (solid), and 
$10^{-4} \ \Zsun$ (dashed), respectively.
Red thick and blue thin curves depict the cases with and without grain growth, respectively.
Triangles are plotted at the states where the fragmentation condition (see text) is met. 
}
\label{fig:nT}
\end{figure}

\begin{figure}
\includegraphics[width=9cm]{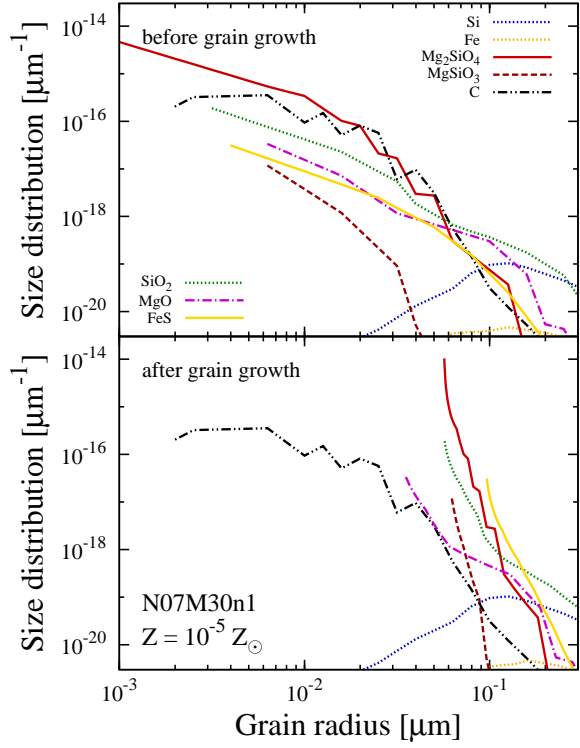}
\caption{
Size distribution function $n_i \varphi_i / \nH$ of dust species $i$
before ($\nH=0.1 \ \percc$) and after ($\nH=10^{16} \ \percc$) grain growth
for our N07M30n1 model with $Z=10^{-5} \ \Zsun$.
Size of all species but for C and Si grains increases by accretion of
heavy elements.
C and Si grains do not grow because the key species, C and Si atoms, are depleted onto
CO and SiO molecules, respectively.
}
\label{fig:dist}
\end{figure}

\begin{figure}
\includegraphics[width=9cm]{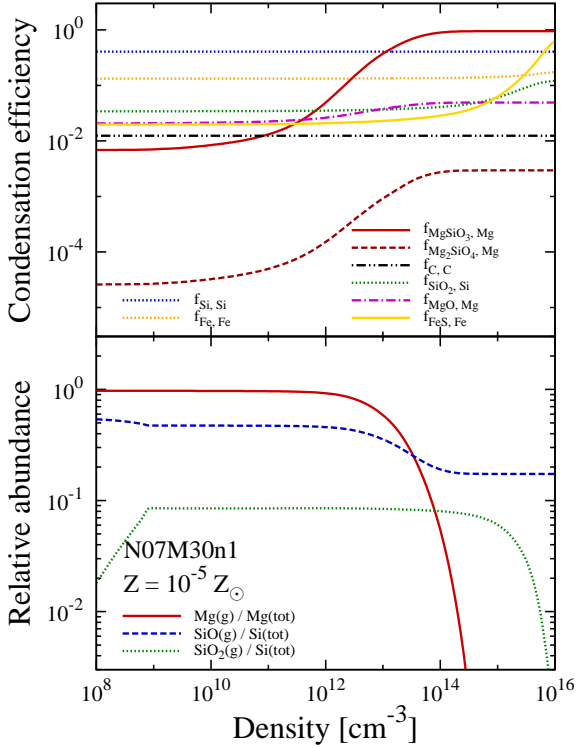}
\caption{
Top: Number fraction of nuclei which are concentrated into grains $i$
relative to total nuclei of element $j$ in both the gas- and solid-phases
(condensation efficiency $f_{ij}$).
Bottom: Number fraction of nuclei which are in the form of
SiO molecules (blue dashed), and SiO$_2$ molecules (green dotted)
relative to total Si nuclei.
Red solid curves depict the number fraction of Mg in the gas phase to total Mg nuclei.
This figure shows the evolution for the same case as Figure \ref{fig:dist}.
}
\label{fig:nf}
\end{figure}

\begin{figure}
\includegraphics[width=9cm]{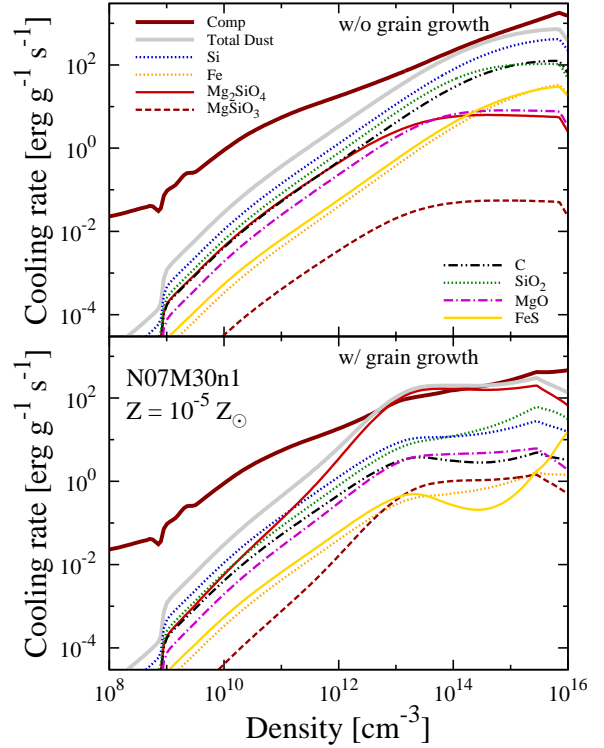}
\caption{
Gas cooling rate per unit gas mass owing to each dust species as a function of gas density
with metallicity $Z=10^{-5} \ \Zsun$ for N07M30n1 models with (top) and without (bottom)
grain growth.
We also plot the total dust cooling rate (thick grey) summed over all grain species and
the gas compressional heating rate (thick dark red). 
}
\label{fig:nLd}
\end{figure}

\begin{table*}
\caption{Critical Conditions with Grain Growth
\label{tab:Crit}}
\begin{tabular}{@{}lclrrrrrrr}
\hline
Model & $\Mpr$ & rev. &
$r_{\rm Sil}^{\rm grow}$ &
log($n_{\rm H, Sil}^{\rm grow}$) &
$f_{\rm dep,Sil,*}$ &
$S_{\rm Sil, *}$ & 
$f_{\rm dep,*}$ &
${\cal D}_{\rm cr,gg}$ & 
$[Z_{\rm cr,gg}]$ \\
\hline 
N07M13n0    & 13 &   n0 & $0.0060$ &       no & $      0.024 $ & $      3.83 $ & $ 0.351$ & $ 0.769$ & $  -5.9$ \\
N07M13n0.1  &    & n0.1 & $0.0444$ &       no & $      0.017 $ & $      2.78 $ & $ 0.270$ & $ 1.473$ & $  -5.5$ \\
N07M13n1    &    &   n1 & $0.0331$ & $  15.2$ & $      0.001 $ & $      3.95 $ & $ 0.128$ & $ 0.928$ & $  -5.3$ \\
N07M13n10   &    &  n10 & $0.0147$ & $  15.9$ & $     <0.001 $ & $      8.45 $ & $ 0.043$ & $ 0.312$ & $  -5.1$ \\
N07M20n0    & 20 &   n0 & $0.0074$ & $ \star$ & ${\bf  0.051}$ & ${\bf  5.22}$ & $ 0.224$ & $ 2.777$ & $  -5.2$ \\
N07M20n0.1  &    & n0.1 & $0.0342$ & $  12.8$ & ${\bf  0.034}$ & ${\bf  3.12}$ & $ 0.146$ & $ 2.184$ & $  -5.1$ \\
N07M20n1    &    &   n1 & $0.0313$ & $  14.0$ & ${\bf  0.003}$ & ${\bf  4.00}$ & $ 0.057$ & $ 1.397$ & $  -4.9$ \\
N07M20n10   &    &  n10 & $0.0228$ & $  15.6$ & ${\bf <0.001}$ & ${\bf  5.25}$ & $ 0.013$ & $ 0.820$ & $  -4.5$ \\
N07M2513n0  & 25 &   n0 & $0.0044$ & $ \star$ & ${\bf  0.060}$ & ${\bf  4.24}$ & $ 0.241$ & $ 3.810$ & $  -5.1$ \\
N07M25n0.1  &    & n0.1 & $0.0615$ & $ \star$ & ${\bf  0.051}$ & ${\bf  1.44}$ & $ 0.200$ & $ 3.103$ & $  -5.1$ \\
N07M25n1    &    &   n1 & $0.0754$ & $  13.8$ & ${\bf  0.020}$ & ${\bf  1.39}$ & $ 0.121$ & $ 2.351$ & $  -5.0$ \\
N07M25n10   &    &  n10 & $0.0511$ & $  15.0$ & ${\bf <0.001}$ & ${\bf  2.93}$ & $ 0.036$ & $ 1.448$ & $  -4.7$ \\
N07M30n0    & 30 &   n0 & $0.0221$ & $ \star$ & ${\bf  0.069}$ & ${\bf  3.99}$ & $ 0.184$ & $ 2.916$ & $  -5.1$ \\
N07M30n0.1  &    & n0.1 & $0.0339$ & $  12.7$ & ${\bf  0.044}$ & ${\bf  4.01}$ & $ 0.128$ & $ 2.265$ & $  -5.0$ \\
N07M30n1    &    &   n1 & $0.0120$ & $  13.2$ & ${\bf  0.006}$ & ${\bf  8.80}$ & $ 0.045$ & $ 0.628$ & $  -5.1$ \\
N07M30n10   &    &  n10 & $0.0246$ & $  15.9$ & $     <0.001 $ & $      5.43 $ & $ 0.015$ & $ 0.734$ & $  -4.6$ \\
N07M170n0   &170 &   n0 & $0.0078$ & $ \star$ & ${\bf  0.072}$ & ${\bf  6.62}$ & $ 0.317$ & $ 3.156$ & $  -5.3$ \\
N07M170n0.1 &    & n0.1 & $0.0439$ & $  13.1$ & ${\bf  0.029}$ & ${\bf  4.07}$ & $ 0.154$ & $ 2.194$ & $  -5.1$ \\
N07M170n1   &    &   n1 & $0.1813$ &       no & ${\bf <0.001}$ & ${\bf  1.17}$ & $ 0.013$ & $ 2.662$ & $  -4.0$ \\
N07M170n10  &    &  n10 &      --- &      --- & $      0.000 $ &           --- & $<0.001$ & $ 1.488$ & $ >-2.0$ \\
N07M200n0   &200 &   n0 & $0.0045$ & $ \star$ & ${\bf  0.053}$ & ${\bf  9.68}$ & $ 0.364$ & $ 3.603$ & $  -5.3$ \\
N07M200n0.1 &    & n0.1 & $0.0452$ & $  13.8$ & $      0.012 $ & $      4.19 $ & $ 0.098$ & $ 1.824$ & $  -5.0$ \\
N07M200n1   &    &   n1 & $0.1440$ &       no & $     <0.001 $ & $      1.48 $ & $ 0.001$ & $ 0.031$ & $  -4.8$ \\
N07M200n10  &    &  n10 &      --- &      --- & $      0.000 $ &           --- & $<0.001$ & $<0.001$ & $  -3.2$ \\
\hline                                                                                         
S12M13norev & 13 & norev& $0.0051$ & $  10.4$ & ${\bf  0.200}$ & ${\bf 25.28}$ & $ 0.602$ & $ 0.826$ & $  -6.0$ \\
S12M13rev1  &    &  rev1& $0.0039$ & $  11.2$ & ${\bf  0.203}$ & ${\bf 14.62}$ & $ 0.476$ & $ 0.497$ & $  -5.9$ \\
S12M13rev2  &    &  rev2& $0.0040$ & $  11.6$ & ${\bf  0.178}$ & ${\bf 10.90}$ & $ 0.333$ & $ 0.336$ & $  -5.8$ \\
S12M13rev3  &    &  rev3& $0.0047$ & $  12.4$ & ${\bf  0.069}$ & ${\bf 10.79}$ & $ 0.164$ & $ 0.188$ & $  -5.7$ \\
S12M20norev & 20 & norev& $0.0159$ & $ \star$ & ${\bf  0.179}$ & ${\bf 14.68}$ & $ 0.370$ & $ 0.466$ & $  -6.2$ \\
S12M20rev1  &    &  rev1& $0.0115$ & $  11.1$ & ${\bf  0.141}$ & ${\bf 11.56}$ & $ 0.242$ & $ 0.204$ & $  -5.9$ \\
S12M20rev2  &    &  rev2& $0.0109$ & $  11.8$ & ${\bf  0.107}$ & ${\bf  9.24}$ & $ 0.172$ & $ 0.118$ & $  -5.7$ \\
S12M20rev3  &    &  rev3& $0.0118$ & $  12.4$ & ${\bf  0.064}$ & ${\bf  7.43}$ & $ 0.105$ & $ 0.069$ & $  -5.5$ \\
S12M25norev & 25 & norev& $0.0080$ & $   9.1$ & ${\bf  0.086}$ & ${\bf 27.19}$ & $ 0.337$ & $ 0.997$ & $  -5.8$ \\
S12M25rev1  &    &  rev1& $0.0061$ & $  11.5$ & ${\bf  0.093}$ & ${\bf 17.77}$ & $ 0.259$ & $ 0.982$ & $  -5.5$ \\
S12M25rev2  &    &  rev2& $0.0060$ & $  12.0$ & ${\bf  0.074}$ & ${\bf 13.58}$ & $ 0.181$ & $ 0.770$ & $  -5.4$ \\
S12M25rev3  &    &  rev3& $0.0061$ & $  12.3$ & ${\bf  0.049}$ & ${\bf 11.21}$ & $ 0.105$ & $ 0.501$ & $  -5.3$ \\
S12M30norev & 30 & norev& $0.0088$ & $ \star$ & ${\bf  0.088}$ & ${\bf 26.22}$ & $ 0.325$ & $ 1.629$ & $  -5.6$ \\
S12M30rev1  &    &  rev1& $0.0065$ & $  11.3$ & ${\bf  0.068}$ & ${\bf 19.56}$ & $ 0.240$ & $ 1.039$ & $  -5.5$ \\
S12M30rev2  &    &  rev2& $0.0062$ & $  12.0$ & ${\bf  0.045}$ & ${\bf 16.09}$ & $ 0.157$ & $ 0.825$ & $  -5.4$ \\
S12M30rev3  &    &  rev3& $0.0071$ & $  12.5$ & ${\bf  0.024}$ & ${\bf 13.24}$ & $ 0.084$ & $ 0.548$ & $  -5.3$ \\
\hline
\end{tabular}
\\ \medskip
Note --- 
$r_i^{\rm grow}$ is the characteristic size of grain species $i$ for grain growth ($\um$; see text).
$n_{{\rm H},i}^{\rm grow}$ is the density where the condensation efficiency of grain species $i$ becomes above
0.5 by grain growth for $Z=10^{-5} \ \Zsun$ runs ($\percc$).
These values are not defined when the condensation efficiency is initially above 0.5 (``$ \star$'')
or when the condensation efficiency do not reach 0.5 until $\nH = 10^{16} \ \percc$ (``no'').
The subscript ``$*$'' denotes the value for $Z=10^{-5} \ \Zsun$ at density $\nH = 10^{12} \ \percc$.
The critical dust-to-gas mass ratio ${\cal D}_{\rm cr,gg}$ and metallicity $Z_{\rm cr,gg}$ are determined
by our one-zone calculations with grain growth.
\end{table*}

\section{Results}

\subsection{Thermal evolution of gas clouds}

We highlight and discuss the results of some specific cases in this section.
Figure \ref{fig:nT} shows the thermal evolution of the cloud center for our N07M30n1 model
with metallicities $Z=10^{-6}$, $10^{-5}$, and $10^{-4} \ \Zsun$.
Around $\nH\sim 10^4 \ \percc$, H$_2$ cooling becomes efficient for a higher metallicity
because H$_2$ molecules are formed on grain surfaces more efficiently.
If the gas temperature drops below $\sim 100$ K, 
HD cooling becomes dominant \citep{Omukai05, Hirano14},
which can be seen at $\nH \sim 10^4$--$10^6 \ \percc $ with metallicities $Z>10^{-5} \ \Zsun $.
Then, OH ($Z\geq 10^{-5} \ \Zsun$) 
becomes a major coolant at $\nH \sim 10^5$--$10^8 \ \percc$.
For $Z=10^{-5}$ and $10^{-4} \ \Zsun$, the gas temperature increases
by the heating owing to the exothermic reaction of the formation of H$_2$ molecules
via rapid three-body reactions at $\nH = 10^{9}$--$10^{12} \ \percc$.
If the amount of dust is sufficiently large, dust cooling becomes effective 
at $\nH = 10^{11}$--$10^{12} \ \percc$.

We examine the condition for the formation of low-mass fragments
by radiative cooling on the basis of the analysis by \citet{Schneider10}.
First, the gas becomes gravitationally unstable and deformed due to
rapid cooling when the specific heat ratio, $\gamma $, drops below unity.
When the gas cooling is inefficient ($\gamma \gtrsim 1$) and 
the gas is likely to collapse in approximately a spherical manner,
yielding fragments whose mass is about the Jeans mass corresponding to
the gas density and temperature at the fragmentation \citep[e.g.][]{Larson85, Inutsuka97}.
Then, the fragmentation condition for a gas cloud is defined as the following set of three criteria:
(1) the gas cloud undergoes rapid radiative cooling ($\gamma < 0.8$), but then
(2) gas cooling becomes soon inefficient ($\gamma > 0.97$).
Since we are interested in the cases where low-mass fragments are formed,
we add another criterion to our fragmentation condition:
(3) the cloud Jeans mass  is less than $0.8 \ \Msun$ when both (1) and (2) are satisfied.

In some models, we find O {\sc i} cooling is efficient enough to trigger
fragmentation, i.e., the criteria (1) and (2) above are met, 
at $\nH \sim 10^2$--$10^3 \ \percc$ for metallicities $Z\geq 10^{-4} \ \Zsun$. 
However, since the mass of the clump is $\sim 100 \ \Msun$ in this regime,
the criterion (3) is not satisfied
%\citet{Ji14} recently suggest that, if the intermediate-mass fragments
%escape from the dense star-forming region, the low-mass star could be
%formed only by the fine-structure cooling. 
%We revisit this in Section 4.
(see Section 4 for the further discussion of this).
It is clearly shown in Figure \ref{fig:nT} that dust cooling is effective at 
higher densities, $\nH = 10^{12}$--$10^{15} \ \percc$,
where the Jeans mass is $\sim 0.01 \ \Msun$.
In this regime, all of the three criteria are met, suggesting that dust cooling 
can drive the fragmentation of the gas into small mass clumps.

Let us discuss the effect of grain growth in detail.
Figure \ref{fig:nT} shows that, for $Z=10^{-5} \ \Zsun $ (solid curves), 
the fragmentation conditions are met with grain growth (red) 
while dust cooling is inefficient without grain growth (blue). 
For $Z=10^{-4} \ \Zsun$ (dashed), the fragmentation conditions are met 
even for the case without grain growth.
Overall, the metallicity, or the initial amount of dust required for the gas fragmentation
is reduced by the effect of grain growth.

Figure \ref{fig:dist} shows the size distribution of grains for N07M30n1 model
with $Z=10^{-5} \ \Zsun$
before ($\nH = 0.1 \ \percc$) and after ($10^{16} \ \percc$) grain growth.
We see that the radii of grains increase especially for $\Olivine $ (red solid curve).
If linearly plotting the figure, the distribution functions
just shift from left to right because the increment of grain
radii is independent from their initial radii as we have mentioned above. 
Figure \ref{fig:nf} shows the condensation efficiencies of grain species $i$ (top) 
and the number fractions of Mg atoms, and SiO and SiO$_2$ molecules 
relative to total Mg and Si nuclei, respectively (bottom).
The condensation efficiencies of all species but carbon (black dot-dot-dashed)
and silicon (blue dotted) increase by accreting the gas-phase species.
Almost all Si atoms are oxidized into SiO molecules at $\nH<10^8 \ \percc$,
where silicate grains ($\Olivine$ and $\Pyroxene$) grow most rapidly by accretion of Mg atoms 
and SiO molecules at $\nH \sim 10^{11}$--$10^{13} \ \percc$.
Further, MgO grains grow almost at the same time as silicate
until gas-phase magnesium is exhausted.
Some fractions of SiO molecules survive without being incorporated into dust grains, because $A_{\Mg} < 2A_{\Si}$.
SiO$_2$ molecules fully condense into $\Silica$ grains at $\nH \gtrsim 10^{15} \ \percc$.
Even if SiO$_2$ molecules are totally depleted,
SiO molecules are not further oxidized because
the major oxidizer, OH, is already depleted into H$_2$O.
At $\nH \sim 10^{15} \ \percc$,
$\Troilite $ grains begin to grow.
The time when grains grow depends mostly on 
the abundances of gas-phase elements as well as 
the monomer radius $a_{ij,0}$.
Carbon grains do not grow because C atoms have been already depleted
into CO molecules at $\nH = 10^6 \ \percc $.
As a result of grain growth,  at $\nH = 10^{12} \ \percc$,
the total depletion factor (total mass fraction of dust relative to metal) 
increases from 0.040 to 0.045 and 0.14
for $Z=10^{-5}$, and $10^{-4} \ \Zsun$, respectively.

Figure \ref{fig:nLd} shows the contribution of each grain species
to gas cooling as a function of gas density for N07M30n1 model.
Without grain growth, Si and $\Silica$ grains make a major contribution to gas cooling.
On the other hand, the cooling rate of $\Olivine$ grains is enhanced 
by accretion of heavy elements, 
and eventually becomes larger than the compressional heating rate. 
This means that the small initial dust amount 
suffices for cloud fragmentation by the effect of grain growth.

\subsection{Conditions for cloud fragmentation}

\subsubsection{Critical dust-to-gas mass ratio and metallicity}

In this section, we discuss the results for all our dust models, and
determine the conditions for the cloud fragmentation.
First of all, we find that dust cooling triggers the gas fragmentation 
for the models where the fragmentation condition is satisfied.
This indicates that the dust-to-gas mass ratio ${\cal D}$ is the key quantity to determine the fragmentation
properties of the gas as suggested by \citet{Schneider12Crit}.
We first show the initial value ${\cal D}_0$ with $Z=10^{-5} \ \Zsun$
in Table \ref{tab:Init_Val}.
The dust-to-gas mass ratio for arbitrary metallicity $Z$ can be obtained simply by 
$ \left( Z/10^{-5} \ \Zsun \right) {\cal D}_0$.
Then, varying the metallicity by 0.1 dex, we define
the critical dust-to-gas mass ratios
${\cal D}_{\rm cr,ng}$ and ${\cal D}_{\rm cr,gg}$ above which dust cooling triggers 
the fragmentation into low-mass gas clumps, based on
our one-zone calculations without and with grain growth
as Tables \ref{tab:Init_Val} and \ref{tab:Crit} show, respectively.
For almost all the models, the critical dust-to-gas mass ratio is reduced by the effect of grain growth.
For example, for N07M30n1 model,  
${\cal D}_{\rm cr,ng} = 6.3\E{-8}$ without grain growth, while
it decreases to ${\cal D}_{\rm cr,gg}=0.63\E{-8}$ with grain growth.
The initial dust-to-gas mass ratio 
${\cal D}_0$ for this case is $0.79\E{-8}$ with $Z=10^{-5} \ \Zsun$,
which is lower than ${\cal D}_{\rm cr,ng}$ but higher than ${\cal D}_{\rm cr,gg}$.
For all our supernova models, the critical dust-to-gas mass ratio is reduced from
${\cal D}_{\rm cr,ng} = [0.81:11.6]\E{-8}$ to ${\cal D}_{\rm cr,gg} = [0.07:3.82]\E{-8}$,
depending on the dust model as we see in the next section.
Note that we do not count n1 and n10 models for PISNe because the PISN models
predict the much larger amounts of heavy elements such as Si and Fe than those
inferred from the Galactic metal-poor stars so far observed, 
and because the ambient gas density of these progenitors
is expected to be small $n_{\rm amb} \lesssim 0.1$--$1 \ \percc$ by the copious emission of ultraviolet
photons in their main sequence \citep{Kitayama04, Whalen04}.

If the dust amount is constant during the collapse of the gas clouds, 
the condition for gas fragmentation can be described simply
by the initial dust-to-gas mass ratio \citep{Schneider12Crit}.
We have shown that grain growth can alter this simple picture.
If grains can completely accrete the gas-phase refractory elements, 
the condition becomes
insensitive to the initial dust properties but sensitive to the gas metallicity
because the depletion efficiency of metals onto dust grains converges to 
a certain value determined by the amounts of the refractory elements in this extreme case.
To see how the fragmentation property depends on the metal and dust contents,
let us study the dependence of the critical metallicity on the dust model.
Although there are various quantities that characterize the dust for each 
supernova model, 
we here focus on the initial depletion factor.
The metallicity $\Zcrit$ above which gas fragmentation is triggered is 
given by the relation $\Zcrit = {\cal D}_{\rm cr} \fdepini ^{-1}$.
The last column in Tables \ref{tab:Init_Val} and \ref{tab:Crit} shows the values
$Z_{\rm cr,ng}$ and $Z_{\rm cr,gg}$ without and with grain growth, respectively,
and Table \ref{tab:Init_Val} shows $\fdepini$.
Without grain growth, the critical metallicity is roughly inversely proportional
to the initial depletion factor as $Z_{\rm cr,ng} = \overline{{\cal D}_{\rm cr,ng}} \fdepini ^{-1}$
with the average critical dust-to-gas mass ratio $\overline{{\cal D}_{\rm cr,ng}} =[2.0:2.5]\E{-8}$.
%With grain growth, we find the following relationship
With grain growth, the least-squares fitting to the results for both 
\Noz \ and \Sch \ models leads to the following relationship: 
\begin{equation}
\left( \frac{Z_{\rm cr,gg} }{ 10^{-5.5} \ \Zsun} \right) = 
\left( \frac{\fdepini }{ 0.18} \right) ^{-0.44 \pm 0.21}.
\label{eq:Zcr}
\end{equation}
%The critical metallicity is still slightly dependent on the initial depletion factor,
%because the gas-phase heavy elements partly condense into dust grains
%as we discuss below. 
The critical metallicity is still dependent on the initial depletion factor, because the gas-phase 
heavy elements partly condense into dust grains as we discuss below.
Note that the uncertainty of the spectrum index stems from the dependence of the critical metallicity 
on the composition, size distribution, and initial condensation efficiency of the dust models adopted
in this paper is discussed in the next subsection.

Our study has shown that the grain growth is important to alter the fragmentation property
for our supernova models.
For the case with the least depletion factor (0.011) among our supernova dust models,
$Z_{\rm cr}$ is reduced by about a factor of 20.
We also find that the dust properties in the early star-forming regions are much 
different from those of the local universe, where all refractory elements 
are depleted onto grains
\citep{Pollack94}.

\begin{figure*}
\includegraphics[width=18cm]{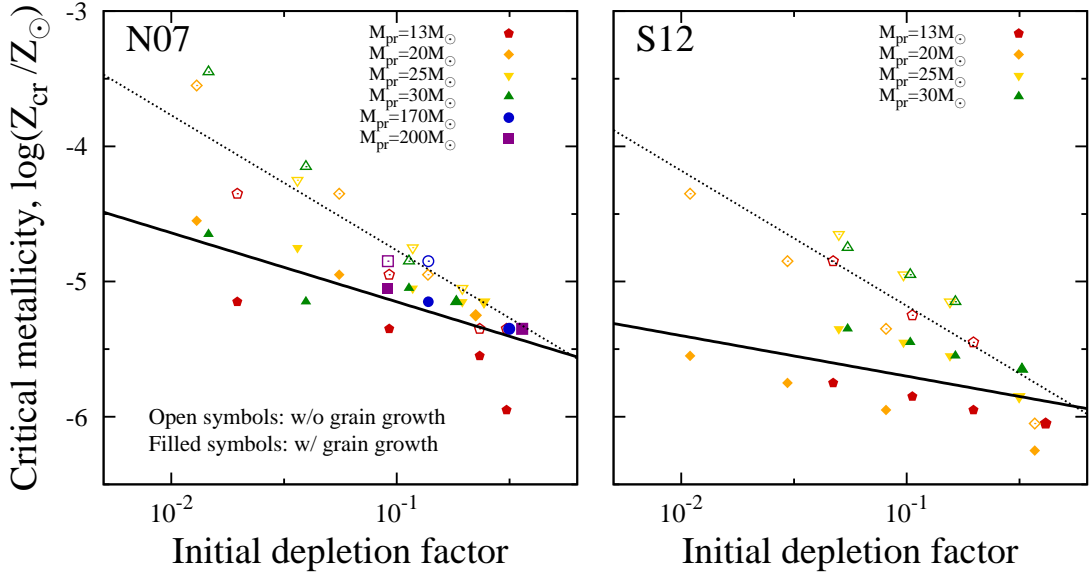}
\caption{
Critical metallicity $\Zcrit$ for \Noz \ (left panel) and \Sch \ (right) supernova dust models
as a function of the initial dust-to-metal mass ratio (depletion factor), $\fdepini$.
Four points with the same symbol (progenitor mass) correspond to our four reverse shock models: 
n0, n0.1, n1, and n10 from right to left.
Open and Filled symbols depict the results obtained by our one-zone calculations with and without
grain growth, respectively.
Dotted and solid lines are fitted to these results for each supernova model.
The former are $Z_{\rm cr,ng} = 3.4\E{-8} \fdepini ^{-1}$ and $Z_{\rm cr,ng} = 1.3\E{-8} \fdepini ^{-1}$, and 
the latter $(Z_{\rm cr,gg} / 10^{-5.5} \ \Zsun) = (\fdepini / 0.50)^{-0.51}$ and
$(Z_{\rm cr,gg} / 10^{-5.5} \ \Zsun) = (\fdepini / 0.022)^{-0.30}$ for \Noz \ and \Sch \ models, respectively.
}
\label{fig:crit}
\end{figure*}

\subsubsection{Dependence of the critical metallicity on the initial dust models}

We note that there are appreciable scatters in the critical metallicity 
from the average value of Equation (\ref{eq:Zcr}),
depending on the initial dust properties.
Figure \ref{fig:crit} shows $\Zcrit$ as a function of 
the initial depletion factor for \Noz \ and \Sch \ models separately.

We begin with the cases without grain growth.
Open symbols in Figure \ref{fig:crit} present $Z_{\rm cr,ng}$, and
the average values are drawn by the dotted lines.
The critical metallicity roughly follows the relationship 
$Z_{{\rm cr,ng}} = {\cal D}_{\rm cr,ng} \fdepini ^{-1}$ 
with the specific value of ${\cal D}_{\rm cr,ng}$ for each progenitor mass.
In more quantitative manner, we revisit the criterion of
\begin{eqnarray}
S{\cal D} &>& 1.4\E{-3} \ {\rm cm^2 \ g^{-1}} \nonumber \\
&\times& \left( \frac{T}{10^3 \ {\rm K}} \right) ^{-1/2}
\left( \frac{\nH}{10^{12} \ \percc} \right) ^{-1/2}, \nonumber
\end{eqnarray}
where $S$ is the total geometrical cross-section of grains per unit 
dust mass \citep{Schneider12Crit}.
We rewrite the left-hand side as  $S{\cal D} = Z \sum _i f_{{\rm dep},i} S_i$
to see the contributions of the dust composition and size distribution of different species.
Here, $f_{{\rm dep},i} = \rho _i / \rho _{\rm metal} = f_{ij} A_j \nH \mu _{ij} \mH / \rho _{\rm metal}$ 
is proportional to the mass fraction of the grain species $i$.
$S_i$ is the cross-section of species $i$ per unit dust mass as
\begin{equation}
S_i 
= \frac{ \pi \left\langle r^2 \right\rangle _i}{(4\pi /3) s_i \left\langle r^3 \right\rangle _i}
= \frac{3}{4 s_i r_i^{\rm cool}},
\label{eq:S_i}
\end{equation}
where 
$r_i^{\rm cool}=\left\langle r^3 \right\rangle _i / \left\langle r^2 \right\rangle _i$ 
is an average dust radius characterizing dust cooling rate.
This represents the contribution of the dust size distribution to gas cooling.

Table \ref{tab:Init_Val} shows $f_{{\rm dep},i}$ and $S_i$
for the major species: carbon and silicate.
For \Sch \ model, the mass fraction of carbon grains is larger than \Noz \ model
as shown in the fourth column of Table \ref{tab:Init_Val}.
Although the mass fraction of silicate grains is similar between the two models,
the contribution of this species to gas cooling is different because the
value $S_{{\rm Sil},0}$ for \Sch \ model is larger than \Noz \ model
as a result of the smaller characteristic size of silicate grains for the former case
as we have seen in Figure \ref{fig:size}.
The contribution of magnetite grains, whose cooling efficiency is also large, can reduce the
critical metallicity for \Sch \ model.
%Instead, for \Noz \ model, silicon grains, whose cooling efficiency is relatively small, accounts for 
%considerable fraction.
Therefore, the total dust cooling efficiency is larger for \Sch \ model
and hence the lower initial dust amount suffices to activate the gas fragmentation.

The critical dust-to-gas mass ratio ${\cal D}_{\rm cr, ng}$ varies also with progenitor masses.
For \Noz \ model,  this value is smallest for M13 models, followed by M25.
This variation stems from the composition of the dominant species: carbon and silicate.
%This can be understood by the dominant contributions of carbon and silicate grains.
For n0 cases, ${\cal D}_{\rm cr, ng}$ is rather insensitive to the progenitor mass
because the sum of the contributions of carbon and silicate is similar to each other.
On the other hand, for n10 cases, the cooling efficiency is largely determined by the mass fraction of carbon grains 
because silicate, which is more efficiently destroyed by the reverse shock than carbon, 
can no longer contribute to gas cooling (see Figure \ref{fig:dust_Nu}).
%In these cases, 
The fraction of carbon grains is largest for M13, and thus
${\cal D}_{\rm cr,ng}$ is smallest, followed by M25.
For \Sch \ model, the critical dust amount is sensitive to the dust size distribution 
of only carbon grains
because this species is the dominant coolant.
${\cal D}_{\rm cr,ng}$ decreases roughly with the increasing 
$S_{\rm C,0}$.

With grain growth, the critical metallicity is reduced, as seen by the filled symbols 
in Figure \ref{fig:crit}.
The solid lines represent the results of the least-squares fitting as
$Z_{\rm cr,gg} \propto \fdepini ^{-0.51}$ and $\fdepini ^{-0.30}$ for \Noz \ and \Sch, respectively,
showing that the effect of grain growth is larger for the latter case.
We compare the density $n_{{\rm H},i}^{{\rm grow}}$
at which the condensation efficiency of grain $i$ exceeds 0.5.
Table \ref{tab:Crit} presents $n_{{\rm H, Sil}}^{{\rm grow}}$ for silicate grains, which grow most rapidly for most of our supernova models.
For \Sch \ model, $n_{{\rm H, Sil}}^{\rm grow}$ is below a density $\nH \sim 10^{12} \ \percc$ where
dust cooling becomes efficient.
On the other hand, for \Noz \ model, grains can grow only after this threshold density.
Thus, the cooling efficiency for \Noz \ model is less modified by grain growth than \Sch \ model
especially for the cases with the dust destruction in the supernovae.

The growth rate of grains is determined by their composition and size.
Since the initial dust composition is similar for \Noz \ and \Sch \ models,
the growth rate is determined largely by the dust size distribution.\footnote{In our previous
paper \citep{Chiaki14}, where we investigate the effect of grain growth on the formation of the specific star \Caf ,
employing the part of \Sch \ model, we conclude that the growth rate is almost insensitive to the initial dust models.
We in this paper survey a wider range of the initial conditions including \Noz \ model, which
predicts larger grain radii than \Sch \ model by about an order of magnitude.
Thus, the dependence of the growth rate on the initial dust size becomes apparent.}
The characteristic density $n_{{\rm H},i}^{\rm grow}$ where grains rapidly grow decreases with 
decreasing dust size because the total cross-section of grains per unit dust mass
is larger for smaller grains. 
Although the relation between $n_{{\rm H},i}^{\rm grow}$ and $f_{ij,0}$ is complicated,
we find a fitting formula to the density where grains rapidly grow as
\[
n_{{\rm H},i}^{\rm grow}
=
1.0\E{12} \ \percc
\left( \frac{A_j}{7.1\E{-10}} \right) ^{-2}
\left( \frac{f_{ij,0}}{0.1} \right) ^{-0.8}  \nonumber
\]
\[
\times 
\left( \frac{r_{i}^{{\rm grow}}}{0.01 \ \um} \right) ^{2}
\left( \frac{a_{i,0}}{1 \ {\rm \AA}} \right) ^{-6}
\left( \frac{m_{i1}}{\mH} \right),
\]
which is valid in the range of $f_{ij,0} \lesssim 0.5$.
In the above equation, $r_i^{{\rm grow}} = \left\langle r^3 \right\rangle ^{1/3}$ 
a measure of the average dust size which characterizes the growth rate.
%This characteristic size contains the information of size distribution.
Table \ref{tab:Crit} shows that $r_i^{{\rm grow}}$ is generally larger for \Noz \ model
as we discuss in Section 2.3.
Thus, the growth rate is smaller for this model.

The critical metallicity with grain growth depends also on the progenitor mass.
For N07M13 model, carbon grains grow because $\rm C>O$.
The condensation efficiency of carbon increases from the initial value 
of $f_{\rm C,C,0} = 0.04$--$0.38$ 
up to $f_{\rm C,C,*} = 0.13$--$0.55$ at $\nH \sim 10^{12} \ \percc$ 
with $Z=10^{-5} \ \Zsun$.
Since the carbon abundance is large relative to magnesium, $\log (A_{\rm C}/A_{\rm Mg})=1.24$,
the growth of carbon grains enhances gas cooling more efficiently than silicate.
The rate of heat transfer between gas and dust becomes proportional to
$f_{\rm dep,C,*} S_{\rm C,*} = [2.6$--$30.3]\E{4}$, 
which is larger than the values for silicate grains with the other progenitor masses.
Therefore, $Z_{\rm cr, gg}$ is smallest for M13.
For N07M13n0 model, $\Magnesia$ grains, which grow at gas density $\log (n_{\rm H, \Magnesia}^{\rm grow}) =13.8$,
further enhance the efficiency of dust cooling.
With the other progenitor masses for \Noz \ model,
silicate grains becomes the dominant species.
For M20 and M30 models, silicate grow too slowly, at $\nH>10^{15} \ \percc$, 
to enhance the cooling efficiency for n10 models.
In these cases, $\Silica $ grains become dominant species for gas cooling.
For M25, the contribution of silicate grains to gas cooling becomes comparable to carbon by
grain growth for n1 and n10 models.
Regardless of the different composition of carbon to silicate, $Z_{\rm cr,gg}$ is
within $\sim 0.2 \ \dex$ from the value for M20 and M30 models.

For S12 model, the dominant species is changed from carbon to silicate by grain growth
for all progenitor masses.
Magnetite grains also contribute to gas cooling for M13 and M30 models via their growth
at densities $\log (n_{\rm H, \Magnetite}^{\rm grow}) = 11.8$--$12.7$ and $12.7$--$13.8$ for the former and latter
models, respectively.
Figure \ref{fig:crit} shows that the critical metallicity is small for M13 and M20 models.
For M13, both magnetite and silicate grow to become dominant.
Although this occurs also for M30 models, the mass density of silicate $f_{\rm dep, Sil, *}$ is larger for M13 as shown in
Table \ref{tab:Crit}.
In addition, both the values $f_{\rm dep, \Magnetite ,*}$ and $S_{\Magnetite ,*}$ are larger for M13 than M30:
$(f_{\rm dep, \Magnetite , *} , S_{\Magnetite , *} )= ( 0.026$--$0.11, 18.0$--$40.9)$ for M13, while 
$(0.004$--$0.070, 15.2$--$19.9)$ for M30.
For M20, the contribution of carbon grains to gas cooling is still large because of 
their small size,
which further reduces the value $Z_{\rm cr,gg}$.

\begin{figure}
\includegraphics[width=9cm]{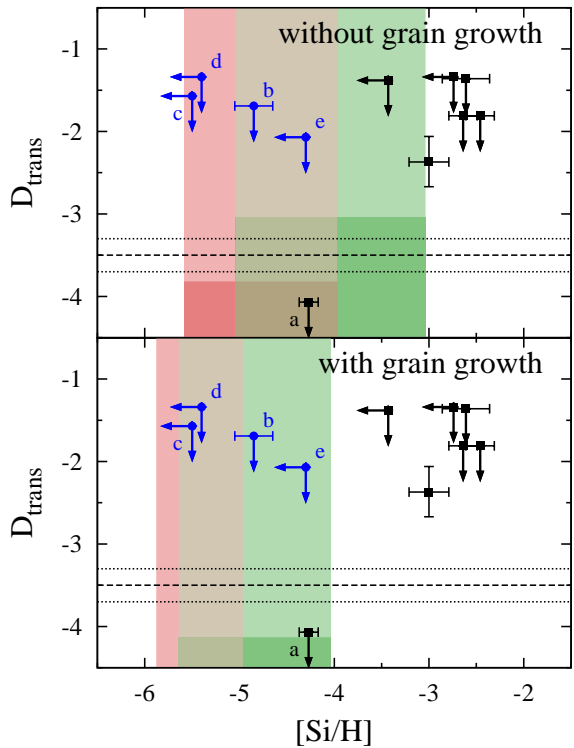}
\caption{
Critical conditions for the formation of low-mass stars plotted on the $D_{\rm trans}$-$\rm [Si/H]$ plane
as used by \citet{Ji14}.
We show the range of the critical metallicity as the green- and red-colored regions defined by
our one-zone calculations with \Noz \ and \Sch \ models, respectively, without (top panel) and
with (bottom) grain growth.
The subsets of the regions with darker colors show the range of C and O abundances examined here.
We also draw horizontal lines depicting the proposed discriminant, $D_{\rm trans} = -3.5 \pm 0.2$,
above which the fine-structure cooling activate the fragmentation \citep{FrebelNorris13}.
We also plot the abundances of extremely metal-poor (EMP: black squares) and
carbon-enhanced metal-poor (CEMP: blue circles) stars which have so far been observed.
The symbols with a--e indicate the elemental abundances of the recently discovered 
primitive star 
\Caf \ \citep[a:][]{Caffau11Nat}, 
and most iron-poor CEMP stars
HE0557-4840 \citep[b:][]{Norris07}, 
HE0107-5240 \citep[c:][]{Christlieb04}, 
HE 1327-2326 \citep[d:][]{Frebel08}, and
SMSS J0313-6708 \citep[e:][]{Keller14}.
The other stars are the carbon-normal main-sequence stars with the intermediate metallicities
taken from the SAGA database \citep{Suda08}.
}
\label{fig:C_Si}
\end{figure}

\section{Conclusion and discussion}

We have investigated the conditions for formation of the first low-mass stars in
the early universe in a low-metallicity gas, by performing one-zone collapse calculations
including grain growth.
As the initial abundances of metal and dust and size distribution of grains,
we have employed the model of dust formation and destruction in preceding Pop III 
supernovae presented by
\Noz \ and \Sch.
Without grain growth, the critical metallicity is inversely proportional to
the initial depletion factor as 
$Z_{\rm cr,ng} = {\cal D}_{\rm cr,ng} \fdepini ^{-1}$ with 
the critical dust-to-gas mass ratio ${\cal D}_{\rm cr,ng} = [0.81:11.6]\E{-8}$,
depending on the supernova models.
${\cal D}_{\rm cr, ng}$ is smaller for \Sch \ model than \Noz \ because, the mass fractions of 
carbon and magnetite grains are larger for the former model, and silicon grains, 
which have less cooling efficiency, account for a considerable mass fraction for \Noz \ model. 
For \Sch \ model, ${\cal D}_{\rm cr, ng}$ is determined mainly by the size of carbon grains,
which are dominant coolants for all progenitor masses.

With grain growth, the critical dust-to-gas mass ratio is reduced down to ${\cal D}_{\rm cr,gg}=[0.07:3.82]\E{-8}$.
The corresponding metallicity is around $Z_{\rm cr,gg} \sim 10^{-5} \ \Zsun$,
proportional to $\fdepini ^{-0.44}$ (Equation (\ref{eq:Zcr})).
This dependence on $\fdepini$ becomes milder than the case without grain growth, but
the dependence is not completely washed out because of the incomplete accretion
of the gas-phase species onto grains.
For \Sch \ model, the accretion of magnesium onto silicate grains rapidly occur.
Carbon atoms are depleted on CO molecules before accreted by carbon grains.
For \Noz \ model, silicate grains do not completely accrete the gas-phase species because
of larger grain radii than \Sch \ model.

It is conceivable that 
dust evaporation and coagulation by grain-grain collision
could reduce the cross-section of collisions between dust grains 
and gas particles.
These processes, however, have little effect on the thermal evolution 
of the clouds in the
region of our interest for the following reasons:
First, silicate grains are sublimated at temperature $\sim 1100 \ {\rm K}$ at high density
\citep{Pollack94}.
Gas temperature rises up to such a value only when the dust cooling is not efficient enough to induce
cloud fragmentation.
Next, grain-grain collision is effective only at densities 
$\nH > 10^{16} (Z/10^{-5} \Zsun)^{-2} (T/1000 \ {\rm K})^{-1} \ \percc$ \citep{Hirashita09}, 
where the gas is already optically thick.
Therefore, we ignore these effects in this study.
We should also mention that, if the sticking coefficient $\alpha _{i}$ is less than unity,
the growth rate of grains is reduced accordingly.
This value is still  uncertain but considered to be 0.1--1 
by various approaches \citep{LeitchDevlin85,Grassi11,Tachibana11}.
We have explicitly tested cases with $\alpha _i = 0.1$
and found that grain growth hardly affects the fragmentation condition:
$(\Zcrit / 10^{-5.5} \ \Zsun) = (\fdepini / 3.5)^{-0.92}$.
In this case, the scaling between $Z_{\rm cr} $ and ${\cal D}_{\rm cr}$ is almost the same
as expected without grain growth.
Let us finally remark that an experimental measurement reveals that the large value of the sticking probability
is order of unity, although this is for Fe grains \citep{Tachibana11}.

As discussed above, the critical metallicity depends on the composition and 
size distribution of dust.
However, it is not possible to observe directly the properties of dust grains 
in the formation site of
long-lived low-metallicity stars discovered in the Galaxy.
Instead, we define the domain of metal abundances where the formation of low-mass 
stars is favored.
We focus here on carbon-normal stars, which are characterized by
$\rm [C/Fe] < 0.7$ \citep{Beers05,Aoki07}.
In such cases, we find that silicate grains are the most important species for 
the gas cooling for the most
cases styled in bold in Tables \ref{tab:Init_Val} and \ref{tab:Crit}. 
We define the range of the critical conditions in terms of silicon abundance
from Equation (\ref{eq:Aj}), setting the metallicity $Z=Z_{\rm cr}$,
which are shown by the green and red shaded regions in Figure \ref{fig:C_Si}, 
for \Noz \ and \Sch \ models, respectively.
Our analyses reveal that the dust-induced fragmentation is activated in the region
inside the color-shaded regions.
Interestingly, the carbon, oxygen and silicon 
abundances of the extremely metal-poor (EMP) 
stars which have so far been observed are located to the right of the color-shaded region.
The formation of the most primitive star ever observed, \Caf \
with $\rm [Si/H]=-4.3$ \citep{Caffau11Nat} is favored by about half of our supernova models even if grain growth 
does not occur.
Furthermore, most of our models can explain the formation of this star by grain growth
\citep{Chiaki14}.

An empirical model for the formation of the first low-mass stars posits that
the fine-structure cooling by carbon and oxygen drives the gas fragmentation
under the critical discriminant of
$D_{\rm trans} = \log (10^{\rm [C/H]} + 0.9 \times 10^{\rm [O/H]}) > -3.5 \pm 0.2$
\citep[horizontal lines in Figure \ref{fig:crit}: e.g.,][]{Bromm01, Frebel05}.
However, this occurs only at the low densities $\nH=10^2$--$10^3 \percc$, where 
the Jeans mass is $ \sim 100 \ \Msun$, which indicates that
the low-mass fragments could not be formed by the fine-structure cooling only
\citep{Omukai05, Schneider06}.
\citet{Ji14} determine the critical Si abundance above which
dust cooling can induce gas fragmentation
by equating the gas compressional heating rate and dust cooling rate,
assuming that dust consists only of Si-bearing grains.
They suggest that the formation of the low-mass stars b--e in Figure \ref{fig:crit}
can not be explained by dust cooling scenario because of the lack of silicon abundance, but can by
fine-structure cooling scenario because of more abundant C and O than their discriminant.
Although one-zone calculations reveal that subsolar-mass clumps could not be formed
in the fine-structure cooling model,
\citet{Ji14} speculate that low-mass star formation could still be possible with help of some 
dynamical effects.
It is important to note that the contribution of carbon grains,
which can reduce the critical Si abundances, 
is uncertain in their model.
The mass ratio of this species is determined for our supernova dust models as
0.003--0.65 and 0.26--0.96 for \Noz \ and \Sch, respectively. 
The contribution of the considerable amount of the carbon grains reduces the 
lower bound of our critical Si abundance
down to the value lower than that of the stars b--e even without grain growth.
Grain growth can further reduce the critical Si abundance
so that a part of \Noz \ models become favored to the formation of these primitive stars. 

We in this work focus on the progenitor models which predict the elemental abundance
consistent with the carbon-normal stars.
The range of $D_{\rm trans}$ is defined from carbon and oxygen abundances
of our progenitor models described as the region with dark color shades in
Figure \ref{fig:C_Si}.
The carbon-enhanced iron-poor stars 
with $\rm [C/Fe] > 0.7$ should lie the upper regime than the dark-colored region,
where our model fails to reproduce the properties of these stars.
To explain the formation path of the CEMP stars,
we would need to consider additional processes such as
the metal pollution by the faint supernovae \citep{Umeda02}, and
the elemental transfer from companion stars \citep{Suda04}.
In \citet{Marassi14}, we discuss the formation of the
most primitive carbon-enhanced star, SMSS J0313-6708, assuming that the 
parent cloud of this star
is polluted with metal and dust by the preceding faint supernova explosions.
In such cases, the carbon is dominant grain species, and
the grain growth is not important because
these grains can hardly accrete carbon atoms because of depletion into CO molecules.
Therefore, the fragmentation property of the gas clouds is determined by
the initial dust-to-gas mass ratio.
Our further comprehensive studies for the faint supernova models, as well as our present
work on carbon-normal stars, can reveal the entire formation processes of the various 
categories of long-lived metal-poor stars.

%%%%%%%%%%%%%%%%%%%%%%%%%%%%%%%%%%%%%%%%%%%%%%
% ACKNOULEDGEMENTS %%%%%%%%%%%%%%%%%%%%%%%%%%%%%%%%
%%%%%%%%%%%%%%%%%%%%%%%%%%%%%%%%%%%%%%%%%%%%%%
\section*{acknowledgments}

We thank Simone Bianchi for his kind contribution.
GC is supported by Research Fellowships of the Japan
Society for the Promotion of Science (JSPS) for Young Scientists.
This work is supported by World Premier International Research Center Initiative (WPI Initiative), 
MEXT, Japan and in part by Grant-in-Aid for Scientific Research 
from the JSPS Promotion of Science (22684004, 23224004, 23540324, 25287040, 25287050, and 26400223).
The research leading to these results has received funding from the European 
Research Council under the European Union's Seventh Framework 
Programme (FP/2007-2013) / ERC Grant Agreement n. 306476.
ML acknowledges the following funding sources:
PRIN INAF 2009 ``Supernova Variety and Nucleosynthesis Yields'', and
PRIN MIUR 2010-2011, project ``The Chemical and dynamical Evolution 
of the Milky Way and Local Group Galaxies'', prot. 2010LY5N2T.


\begin{thebibliography}{}

% A %
\bibitem[Abel et al.(2002)]{Abel02} Abel, T., Bryan, G.~L., 
\& Norman, M.~L.\ 2002, Science, 295, 93 
\bibitem[Aoki et al.(2007)]{Aoki07} Aoki, W., Beers, T.~C., 
Christlieb, N., et al.\ 2007, \apj, 655, 492 

% B %
\bibitem[Beers 
\& Christlieb(2005)]{Beers05} Beers, T.~C., \& Christlieb, N.\ 2005, \araa, 43, 531 
\bibitem[Bianchi \& Schneider(2007)]{Bianchi07} Bianchi, S., 
\& Schneider, R.\ 2007, \mnras, 378, 973 
\bibitem[Bromm et al.(2001)]{Bromm01} Bromm, V., Ferrara, A., 
Coppi, P.~S., \& Larson, R.~B.\ 2001, \mnras, 328, 969 
\bibitem[Bromm \& Loeb(2003)]{BrommLoeb03} Bromm, V., \& Loeb, A.\ 2003, Nat, 425, 812 

% C %
%\bibitem[Caffau et al.(2011)]{Caffau11SolPhys} Caffau, E., Ludwig, 
H.~G., Steffen, M., Freytag, B., \& Bonifacio, P.\ 2011, \solphys, 268, 255 
\bibitem[Caffau et al.(2011)]{Caffau11Nat} Caffau, E., Bonifacio, 
P., Fran{\c c}ois, P., et al.\ 2011, \nat, 477, 67 
%\bibitem[Caffau et al.(2012)]{Caffau12} Caffau, E., Bonifacio, P., Fran{\c c}ois, 
%P., et al.\ 2012, \aap, 542, A51 
%\bibitem[Cazaux \& Spaans(2009)]{CazauxSpaans09} Cazaux, S., \& Spaans, 
%M.\ 2009, \aap, 496, 365 
\bibitem[Cazaux 
\& Tielens(2002)]{CazauxTielens02} Cazaux, S., \& Tielens, A.~G.~G.~M.\ 2002, \apjl, 575, L29 
%\bibitem[Chiaki et al.(2013a)]{Chiaki13SN} Chiaki, G., Yoshida, N., 
%\& Kitayama, T.\ 2013, \apj, 762, 50 
\bibitem[Chiaki et al.(2013)]{Chiaki13Single} Chiaki, G., Nozawa, T., 
\& Yoshida, N.\ 2013, \apjl, 765, L3 
\bibitem[Chiaki et al.(2014)]{Chiaki14} Chiaki, G., Schneider, 
R., Nozawa, T., et al.\ 2014, \mnras, 341 
%\bibitem[Clark et al.(2011)]{Clark11} Clark, P.~C., Glover, 
%S.~C.~O., Smith, R.~J., et al.\ 2011, Science, 331, 1040 
\bibitem[Christlieb et al.(2004)]{Christlieb04} Christlieb, N., 
Gustafsson, B., Korn, A.~J., et al.\ 2004, \apj, 603, 708 




% D %
\bibitem[De Cia et al.(2013)]{DeCia13} De Cia, A., Ledoux, C., 
Savaglio, S., Schady, P., \& Vreeswijk, P.~M.\ 2013, \aap, 560, A88 
%\bibitem[Di Criscienzo et al.(2013)]{DiCriscienzo13} Di Criscienzo, 
%M., Dell'Agli, F., Ventura, P., et al.\ 2013, \mnras, 433, 313 
%\bibitem[Dopcke et al.(2011)]{Dopcke11} Dopcke, G., Glover, 
%S.~C.~O., Clark, P.~C., \& Klessen, R.~S.\ 2011, \apjl, 729, L3 
%\bibitem[Dopcke et al.(2013)]{Dopcke13} Dopcke, G., Glover, 
%S.~C.~O., Clark, P.~C., \& Klessen, R.~S.\ 2013, \apj, 766, 103
%\bibitem[Dwek et al.(2007)]{Dwek07} Dwek, E., Galliano, F., 
%\& Jones, A.~P.\ 2007, ApJ, 662, 927 

% F %
\bibitem[Frebel et al.(2005)]{Frebel05} Frebel, A., Aoki, W., 
Christlieb, N., et al.\ 2005, Nat, 434, 871 
\bibitem[Frebel et al.(2007)]{Frebel07} Frebel, A., Johnson, 
J.~L., \& Bromm, V.\ 2007, MNRAS, 380, L40 
\bibitem[Frebel et al.(2008)]{Frebel08} Frebel, A., Collet, R., 
Eriksson, K., Christlieb, N., \& Aoki, W.\ 2008, \apj, 684, 588 
\bibitem[Frebel \& Norris(2013)]{FrebelNorris13} 
Frebel, A., \& Norris, J.~E.\ 2013, Planets, Stars and Stellar Systems.~Volume 5: Galactic Structure and Stellar Populations, 55 

% G %
\bibitem[Grassi et 
al.(2011)]{Grassi11} Grassi, T., Krstic, P., Merlin, E., et al.\ 2011, \aap, 533, A123 
%\bibitem[Greif et al.(2012)]{Greif12} Greif, T.~H., Bromm, V., 
%Clark, P.~C., et al.\ 2012, \mnras, 424, 399 

% H %
\bibitem[Hartquist et al.(1980)]{Hartquist80} Hartquist, T.~W., 
Dalgarno, A., \& Oppenheimer, M.\ 1980, \apj, 236, 182 
\bibitem[Hirashita \& Omukai(2009)]{Hirashita09} Hirashita, H., \& Omukai, K.\ 2009, 
\mnras, 399, 1795 
\bibitem[Hosokawa et al.(2011)]{Hosokawa11} Hosokawa, T., Omukai, 
K., Yoshida, N., \& Yorke, H.~W.\ 2011, Science, 334, 1250 
\bibitem[Hirano et al.(2014)]{Hirano14} Hirano, S., Hosokawa, 
T., Yoshida, N., et al.\ 2014, \apj, 781, 60 

% I %
\bibitem[Inutsuka 
\& Miyama(1997)]{Inutsuka97} Inutsuka, S.-I., \& Miyama, S.~M.\ 1997, \apj, 480, 681 


% J %
\bibitem[Ji et al.(2014)]{Ji14} Ji, A.~P., Frebel, A., 
\& Bromm, V.\ 2014, \apj, 782, 95 

% K %
\bibitem[Keller et al.(2014)]{Keller14} Keller, S.~C., Bessell, 
M.~S., Frebel, A., et al.\ 2014, \nat, 506, 463 
\bibitem[Kitayama et al.(2004)]{Kitayama04} Kitayama, T., Yoshida, 
N., Susa, H., \& Umemura, M.\ 2004, \apj, 613, 631 
%\bibitem[Kitayama 
%\& Yoshida(2005)]{Kitayama05} Kitayama, T., \& Yoshida, N.\ 2005, \apj, 630, 675
\bibitem[Kroupa(2002)]{Kroupa02} Kroupa, P.\ 2002, Science, 295, 
82 

% L %
\bibitem[Langer 
\& Glassgold(1990)]{Langer90} Langer, W.~D., \& Glassgold, A.~E.\ 1990, ApJ, 352, 123 
%\bibitem[Larson(1978)]{Larson78} Larson, R.~B.\ 1978, \mnras, 
%184, 69 
\bibitem[Larson(1985)]{Larson85} Larson, R.~B.\ 1985, \mnras, 
214, 379 
%\bibitem[Larson(2005)]{Larson05} Larson, R.~B.\ 2005, \mnras, 
%359, 211 
\bibitem[Leitch-Devlin 
\& Williams(1985)]{LeitchDevlin85} Leitch-Devlin, M.~A., \& Williams, D.~A.\ 1985, \mnras, 213, 295 
%\bibitem[Li et al.(2003)]{Li03} Li, Y., Klessen, R.~S., 
%\& Mac Low, M.-M.\ 2003, \apj, 592, 975 
\bibitem[Limongi \& Chieffi(2012)]{Limongi12}
Limongi, M., \& Chieffi, A. 2012, ApJS, 199, 38 




% M %
%\bibitem[Machida et al.(2005)]{Machida05} Machida, M.~N., 
%Tomisaka, K., Nakamura, F., \& Fujimoto, M.~Y.\ 2005, \apj, 622, 39 
\bibitem[Marassi et al.(2014)]{Marassi14} Marassi, S., Chiaki, 
G., Schneider, R., et al.\ 2014, \apj, 794, 100 
\bibitem[Mayer 
\& Duschl(2005)]{Mayer05} Mayer, M., \& Duschl, W.~J.\ 2005, \mnras, 358, 614 
\bibitem[Molaro et al.(2000)]{Molaro00} Molaro, P., Bonifacio, 
P., Centuri{\'o}n, M., et al.\ 2000, \apj, 541, 54 

% N %
\bibitem[Norris et al.(2007)]{Norris07} Norris, J.~E., 
Christlieb, N., Korn, A.~J., et al.\ 2007, \apj, 670, 774 
\bibitem[Nozawa et al.(2003)]{Nozawa03} Nozawa, T., Kozasa, T., 
Umeda, H., Maeda, K., \& Nomoto, K.\ 2003, \apj, 598, 785 
%\bibitem[Nozawa et al.(2006)]{Nozawa06} Nozawa, T., Kozasa, T., 
%\& Habe, A.\ 2006, \apj, 648, 435 
\bibitem[Nozawa et al.(2007)]{Nozawa07} Nozawa, T., Kozasa, T., 
Habe, A., et al.\ 2007, \apj, 666, 955 
\bibitem[Nozawa et al.(2008)]{Nozawa08} Nozawa, T., Kozasa, T., 
Tominaga, N., et al.\ 2008, \apj, 684, 1343 
\bibitem[Nozawa et al.(2012)]{Nozawa12} Nozawa, T., Kozasa, T., 
\& Nomoto, K.\ 2012, \apjl, 756, L35 
%\bibitem[Nozawa 
%\& Kozasa(2013)]{Nozawa13} Nozawa, T., \& Kozasa, T.\ 2013, \apj, 776, 24 

% O %
%\bibitem[Omukai \& Nishi(1998)]{OmukaiNishi98} Omukai, K., \& Nishi, R.\ 1998, 
%\apj, 508, 141 
\bibitem[Omukai(2000)]{Omukai00} Omukai, K.\ 2000, \apj, 534, 
809 
\bibitem[Omukai 
\& Palla(2003)]{Omukai03} Omukai, K., \& Palla, F.\ 2003, \apj, 589, 677 
\bibitem[Omukai et al.(2005)]{Omukai05} Omukai, K., Tsuribe, T., 
Schneider, R., \& Ferrara, A.\ 2005, \apj, 626, 627 
%\bibitem[Omukai et al.(2008)]{Omukai08} Omukai, K., Schneider, 
%R., \& Haiman, Z.\ 2008, \apj, 686, 801 
\bibitem[Omukai et al.(2010)]{Omukai10} Omukai, K., Hosokawa, 
T., \& Yoshida, N.\ 2010, \apj, 722, 1793 

% P %
\bibitem[Pollack et al.(1994)]{Pollack94} Pollack, J.~B., 
Hollenbach, D., Beckwith, S., et al.\ 1994, \apj, 421, 615 


%  R %
%\bibitem[Ritter et al.(2012)]{Ritter12} Ritter, J.~S., 
%Safranek-Shrader, C., Gnat, O., Milosavljevi{\'c}, M., 
%\& Bromm, V.\ 2012, \apj, 761, 56 

% S %
\bibitem[Santoro 
\& Shull(2006)]{SantoroShull06} Santoro, F., \& Shull, J.~M.\ 2006, \apj, 643, 26 
\bibitem[Schneider et al.(2003)]{Schneider03} Schneider, R., 
Ferrara, A., Salvaterra, R., Omukai, K., \& Bromm, V.\ 2003, \nat, 422, 869 
\bibitem[Schneider et al.(2006)]{Schneider06} Schneider, R., 
Omukai, K., Inoue, A.~K., \& Ferrara, A.\ 2006, \mnras, 369, 1437 
\bibitem[Schneider 
\& Omukai(2010)]{Schneider10} Schneider, R., \& Omukai, K.\ 2010, \mnras, 402, 429 
\bibitem[Schneider et al.(2012a)]{Schneider12Crit} Schneider, R., 
Omukai, K., Bianchi, S., \& Valiante, R.\ 2012a, \mnras, 419, 1566 
\bibitem[Schneider et al.(2012b)]{Schneider12Caf} Schneider, R., 
Omukai, K., Limongi, M., et al.\ 2012b, \mnras, 423, L60 
\bibitem[Silvia et al.(2010)]{Silvia10} Silvia, D.~W., Smith, 
B.~D., \& Shull, J.~M.\ 2010, \apj, 715, 1575
\bibitem[Silvia et al.(2012)]{Silvia12} Silvia, D.~W., Smith, 
B.~D., \& Shull, J.~M.\ 2012, \apj, 748, 12 
\bibitem[Suda et al.(2004)]{Suda04} Suda, T., Aikawa, M., 
Machida, M.~N., Fujimoto, M.~Y., \& Iben, I., Jr.\ 2004, \apj, 611, 476 
\bibitem[Suda et al.(2008)]{Suda08} Suda, T., Katsuta, Y., 
Yamada, S., et al.\ 2008, \pasj, 60, 1159 
\bibitem[Susa et al.(2014)]{Susa14} Susa, H., Hasegawa, K., 
\& Tominaga, N.\ 2014, \apj, 792, 32 

% T %
\bibitem[Tachibana et al.(2011)]{Tachibana11} Tachibana, S., 
Nagahara, H., Ozawa, K., et al.\ 2011, \apj, 736, 16 
\bibitem[Todini \& Ferrara(2001)]{Todini01} Todini, P., \& Ferrara, A.\ 2001, \mnras, 325, 726 
%\bibitem[Tominaga et al.(2013) in prep.]{Tominaga12} Tominaga, N. et al.\ 2013, in prep.
%\bibitem[Turk et al.(2009)]{Turk09} Turk, M.~J., Abel, T., 
%\& O'Shea, B.\ 2009, Science, 325, 601 
%\bibitem[Tsuribe 
%\& Omukai(2008)]{TsuribeOmukai08} Tsuribe, T., \& Omukai, K.\ 2008, \apjl, 676, L45 

% U % 
\bibitem[Umeda 
\& Nomoto(2002)]{Umeda02} Umeda, H., \& Nomoto, K.\ 2002, \apj, 565, 385 
 

% V %
%\bibitem[Ventura et al.(2011)]{Ventura11} Ventura, P., Di 
%Criscienzo, M., Schneider, R., et al.\ 2011, arXiv:1111.2053 

% W %
\bibitem[Whalen et al.(2004)]{Whalen04} Whalen, D., Abel, T., 
\& Norman, M.~L.\ 2004, \apj, 610, 14 
%\bibitem[Whalen et al.(2008)]{Whalen08} Whalen, D., van Veelen, 
%B., O'Shea, B.~W., \& Norman, M.~L.\ 2008, \apj, 682, 49 
%\bibitem[Wise et al.(2012)]{Wise12} Wise, J.~H., Turk, M.~J., 
%Norman, M.~L., \& Abel, T.\ 2012, \apj, 745, 50 

% X %
%\bibitem[Xu et al.(2013)]{Xu13} Xu, H., Wise, J.~H., 
%\& Norman, M.~L.\ 2013, \apj, 773, 83 

% Y %
\bibitem[Yoshida et al.(2006)]{Yoshida06} Yoshida, N., Omukai, 
K., Hernquist, L., \& Abel, T.\ 2006, \apj, 652, 6 
%\bibitem[Yasuda 
%\& Kozasa(2012)]{Yasuda12} Yasuda, Y., \& Kozasa, T.\ 2012, \apj, 745, 159 


\end{thebibliography}
\end{document}